\documentclass[journal,onecolumn]{IEEEtran}
\usepackage{amsmath,amsfonts,amssymb,amsthm}
\usepackage{array}
\usepackage[caption=false,font=normalsize,labelfont=sf,textfont=sf]{subfig}
\usepackage{textcomp}
\usepackage{stfloats}
\usepackage{url}
\usepackage{verbatim}
\usepackage{graphicx}
\usepackage{cite}
\usepackage{booktabs}
\usepackage[ruled]{algorithm2e}

\newtheorem{lemma}{Lemma}

\newtheorem{remark}{Remark}

\begin{document}

\title{An Extended Object PMB Filter with Zero-Inflated Poisson Measurement Model Using Belief Propagation}

\author{Xueqi Qiu, Yuxuan Xia, Hyowon Kim, Chaoqun Yang
\thanks{X. Qiu is with the Department of Computer Science, Durham University, Durham DH13LT, United Kingdom. Email: xueqi.qiu@durham.ac.uk. Y. Xia is with the School of Automation and Intelligent Sensing, Shanghai Jiao Tong University, Shanghai 200240, China. Email: yuxuan.xia@sjtu.edu.cn. H. Kim is with the Department of Electronics Engineering,
Chungnam National University, 
Daejeon 34134, South Korea. Email: hyowon.kim@cnu.ac.kr. C. Yang is with the School of Automation, Southeast University, Nanjing 210096, China Email: ycq@seu.edu.cn}
}



\maketitle

\begin{abstract}
This paper presents an efficient implementation of the extended object Poisson multi-Bernoulli (PMB) filter under the zero-inflated Poisson (ZIP) object measurement model using particle belief propagation (BP). The ZIP measurement model separates a Bernoulli object detection event from the conditional Poisson generation of object measurements, enabling principled handling of empty measurement sets. Building upon the PMB mixture posterior, we present a factorized joint posterior over set of objects with object detection variables and a dual representation of data association using both object-oriented and measurement-oriented association variables. Notably, this representation replaces the implicit high-order global hypothesis constraint by local consistency factors, yielding a factor graph amenable to BP. In addition, we present a particle-based implementation, where the single object densities of Bernoulli components are represented using particles. Simulation results show that the proposed method achieves filtering performance comparable to a sampling-based PMBM implementation, while having lower runtime. We also validate the efficacy of the proposed method using real-world lidar data for pedestrian tracking.
\end{abstract}


\section{Introduction}

Multi-object tracking (MOT) considers estimating the states of multiple objects from sequences of noisy sensor measurements \cite{bar2011tracking}. High-resolution sensors, such as lidar and automotive radar, often produce multiple measurements from a single object. Tracking such objects is called extended object tracking (EOT), where the object state includes both kinematic and extent parameters \cite{extendedoverview}. In this paper, we focus on multiple EOT with an unknown and time-varying number of objects.

For EOT, the set of single object measurements is commonly modeled using an inhomogeneous Poisson point process (PPP) \cite{ppp}, where the number of measurements is Poisson distributed and the measurements are independently and identically distributed conditioned on the object state. Because of this, the PPP measurement model avoids explicit associations between measurements and reflection points on objects, thereby making it convenient to use in EOT. 

The zero-inflated Poisson (ZIP) model augments the PPP with an explicit Bernoulli detection event. Under this model, an object is first detected with some probability, and only if it is detected, it then generates a Poisson set of measurements. Consequently, unlike the standard PPP model, the ZIP model can account separately for object missed detections and for the Poisson variability in the number of generated measurements. This separation is particularly beneficial for multiple EOT in scenarios with closely-spaced or partially occluded objects, since it provides a more flexible and interpretable measurement model. Both Poisson and ZIP measurement models have been used in many different multiple EOT algorithms, including, e.g., the multiple hypothesis tracker (MHT) \cite{coraluppi2018multiple,tang2019seamless}, the joint probabilistic data association filter \cite{vivone2016joint,yang2020marginal}, and many filters based on random finite sets (RFS) \cite{phdextended2,cphdextended,lmbextended,pmbmextended2,xia2021poisson}.  

For standard multi-object models with Poisson birth and the general measurement model that each object generates an independent set of measurements, the multi-object posterior is of the Poisson multi-Bernoulli mixture (PMBM) form \cite{pmbmextended2,garcia2021poisson,garcia2023poisson}. The PMBM density admits a compact representation of global hypotheses with probabilistic object existence encoded in each Bernoulli component, and it also enables measurement-driven track initiation by representing undetected objects as a PPP \cite{pmbmpoint2}. The PMBM filtering recursions have been derived for point objects \cite{pmbmpoint2}, extended objects \cite{pmbmextended2} and more general measurement models \cite{garcia2021poisson,garcia2023poisson}. The Poisson multi-Bernoulli (PMB) filter is an efficient approximation of the PMBM filter that approximates the PMBM posterior as a PMB \cite{pmbmpoint,xia2021poisson}. The extended object PMBM and PMB filters have been shown to achieve strong performance compared to many other multi-object filters \cite{pmbmextended2,xia2021poisson,xia2023trajectory}.

A central challenge in multiple EOT is the data association problem, which needs to be efficiently addressed to keep the computational complexity of multiple EOT methods tractable. The most common solution is by enumerating data association hypotheses with high weights and by pruning those with negligible weights. Approaches of this kind include the two-step method based on clustering and assignment (C\&A) \cite{lmbextended,pmbmextended2} and single-step sampling-based methods \cite{soextended}. However, the performance of C\&A-based methods often sharply drop in scenarios with objects moving in proximity, and sampling-based methods often have a high computational cost.

A more scalable approach that avoids explicit enumeration of data associations is to directly compute the marginal multi-object posterior density, where the data association uncertainties are marginalized out. This can be achieved by performing belief propagation (BP) on a factor graph representation of the joint posterior over data association variables and multi-object states \cite{florian2021scalable,xia2023trajectory}. In particular, \cite{xia2023trajectory} shows that, for the extended object PMB filter, its particle BP-based implementation outperforms both C\&A-based and sampling-based implementations in a challenging scenario with multiple objects moving in proximity.

However, BP-based implementations in \cite{florian2021scalable,xia2023trajectory} are developed for the PPP measurement model and do not explicitly account for the object detection probability, while the C\&A-based and sampling-based implementations in \cite{pmbmextended2,soextended} can handle the more general ZIP measurement model. The ZIP model makes it more challenging for BP because, when no measurements are generated, this event may result either from a missed object detection under the Bernoulli detection mechanism or from an object being detected but still producing zero Poisson measurements. This ambiguity introduces an additional coupling in the joint posterior, leading to a high-order factor in the corresponding factor graph, which is not well suited for efficient BP.

An analytical BP implementation for the ZIP measurement model was proposed in \cite{li2022loopy} for the gamma Gaussian inverse-Wishart (GGIW) model and later extended to multi-sensor fusion in \cite{li2024multisensor}, but it has neglected the possibility that the Poisson detection branch of the ZIP model might still generate zero measurements. Furthermore, analytical GGIW implementations typically rely on clustering methods to enumerate data association hypotheses in computing the extrinsic messages for Bernoulli components, which can degrade performance in scenarios with closely-spaced objects. This has been improved in \cite{lyu2026fast}, where the intractable terms involved in computing the extrinsic messages are approximately computed using importance sampling. However, the legacy-object factorization in \cite[Eq.~10]{lyu2026fast}, as printed, is not algebraically equivalent to the underlying joint posterior in general; see Appendix~\ref{app:factorization_consistency}. An alternative factor graph formulation under the ZIP model is presented in \cite{ma2024max}, and the max-sum message passing is employed to compute the maximum a posteriori (MAP) data association. Nevertheless, retaining only the MAP solution may result in degraded performance when data association uncertainty is high. Moreover, the corresponding factor for ZIP in the factor graph remains high-order in \cite{ma2024max}, making it unsuitable for efficient BP.

In light of the above discussion, there is a need for an efficient BP-based implementation of the extended object PMB filter under the ZIP object measurement model that can well handle the coupling introduced by the ZIP model and avoid explicit enumeration of data association hypotheses. To this end, we present an efficient implementation of the extended object PMB filter under the ZIP measurement model using BP. We note that in this paper we focus on the estimation of the current object states, and the explicit estimation of object trajectories can be achieved by considering sets of trajectories \cite{garcia2020trajectory,granstrom2024poisson,xia2023trajectory}.

The main contributions of this paper are summarized as follows:
\begin{itemize}
  \item We present a joint factorized posterior for the extended object PMB update under the ZIP measurement model. Specifically, by introducing auxiliary object detection variables and a dual data association representation using both object-oriented and measurement-oriented association variables, we obtain a factor graph amenable to BP.
  \item We derive the corresponding message passing and belief calculation equations for the Bernoulli components.
  \item We develop a particle-based implementation of the proposed extended object PMB filter using BP.
  \item We compare the proposed method with existing sampling-based implementations \cite{soextended} under the ZIP model. The results show that the proposed method achieves filtering performance comparable to a sampling-based PMBM implementation, while having lower runtime.
  \item We validate the efficacy of the proposed method using real-world lidar data for pedestrian tracking.
\end{itemize}

The rest of the paper is organized as follows. In Section \ref{sec_background}, we present the standard multi-object models, the PMBM density and its PMB approximation on an augmented space with auxiliary variables, and the PMB filtering recursion. In Section \ref{sec_joint_posterior}, we present the joint posterior representation for the PMB update with the ZIP object measurement model, and its factorization. In Section \ref{sec_lbp}, we present the equations for performing loopy belief propagation on the factor graph representation of the joint posterior, and in Section \ref{sec_particle} we present the particle-based implementation. In Section \ref{sec_simulation}, we present the simulation results comparing the proposed method with existing methods. Finally, we conclude the paper in Section \ref{sec_conclusion}.

\section{Background}
\label{sec_background}
In this section, we first introduce the standard multi-object dynamic and measurement models with Poisson birth and ZIP object-generated measurements. Next, we present the PMBM density and its PMB approximation on a space with auxiliary variables. Then, we present the PMB filtering recursion.

For a generic space $D$, we denote $\mathcal{F}(D)$ for the collection of all finite subsets of $D$, and for any $A \in \mathcal{F}(D)$, its cardinality is denoted by $|A|$. The symbol $\uplus$ denotes union over mutually disjoint sets, and we write $\langle f,g \rangle$ the inner product $\int f(x)g(x)\,dx$. Given a real-valued function $f(\cdot)$, we use the multi-object exponential representation $f^A$ to denote $\prod_{x\in A} f(x)$, with the convention that $f^\emptyset = 1$. Lastly, the Kronecker delta centered at $x$ is denoted $\delta_x[\cdot]$.

\subsection{Multi-Object Models}
\label{sec_model}

The single object state $x_k \in \mathbb{R}^{d_x}$ at time step $k$ contains information of the object, including e.g., its kinematic state and its extent state, which describes its shape and size. The set of object states at time step $k$ is $\mathbf{x}_k = \{x_k^1,\dots,x_k^{n_k}\} \in \mathcal{F}(\mathbb{R}^{d_x})$, and the set of measurements at time step $k$ is $\mathbf{z}_k = \{z_k^1,\dots,z_k^{m_k}\} \in \mathcal{F}(\mathbb{R}^{d_z})$. The sequence of measurement sets up to and including time $k$ is denoted as $\mathbf{z}_{1:k}$.

\subsubsection{Dynamic model}
At time step $k$, new objects are born following a PPP with birth intensity $\lambda_k^B(x_k)$, independently of any existing objects. Each existing object $x_{k-1}$ survives with probability $p_k^S(x_{k-1})$, and if it survives, it evolves according to the Markovian transition density $g_k(x_k|x_{k-1})$, independently of any other objects. 
\subsubsection{Measurement model}
The set $\mathbf{z}_k$ of measurements is the union of object-generated measurements and clutter measurements. The set of clutter measurements is a PPP with Poisson intensity $\lambda^C_k(z_k) = \gamma^C_k\mu^C_k(z_k)$, where $\gamma^C_k$ is the Poisson rate and $\mu^C_k(z_k)$ is the clutter spatial distribution. 

A single object with state $x_k$ generates an independent set $\mathbf{w}_k$ of measurements with density $\ell_k(\mathbf{w}_k|x_k)$. In this work, we assume that $\mathbf{w}_k$ is a ZIP process, and its set density is 
\begin{align}\label{eq_ppp_meas}
    \ell_k(\mathbf{w}_k|x_k) = \begin{cases}
      p^D_k(x_k)e^{-\gamma_k(x_k)}\left[\gamma_k(x_k)\ell_k(\cdot|x_k)\right]^{\mathbf{w}_k} & \mathbf{w}_k \neq \emptyset \\
      1 - p^D_k(x_k) + p^D_k(x_k)e^{-\gamma_k(x_k)} & \mathbf{w}_k = \emptyset,
    \end{cases}
\end{align}
where $p^D_k(x_k)$ denotes the detection probability. If object $x_k$ is detected, it generates a Poisson set of measurements with rate $\gamma_k(x_k)$, and $\ell_k(\cdot|x_k)$ is the single measurement likelihood. The probability that object $x_k$ generates at least one measurement is given by $1-\ell_k(\emptyset|x_k) = p^D_k(x_k)(1-e^{-\gamma_k(x_k)})$. 

\subsection{PMBM Density}
Given a sequence of measurements $\mathbf{z}_{1:k^\prime}$ up to and including time step $k^\prime \in \{k-1,k\}$ and the multi-object models described in Section \ref{sec_model}, the exact multi-object density at time step $k$ is of the PMBM form with \cite{pmbmextended2}
\begin{align}
    f_{k|k^\prime}\left(\mathbf{x}_k\right) &=  \sum_{\mathbf{x}^u_k \uplus \mathbf{x}^d_k = \mathbf{x}_k}f_{k|k^\prime}^p\left(\mathbf{x}_k^u\right)f_{k|k^\prime}^{mbm}\left(\mathbf{x}_k^d\right),\label{eq_pmbm}\\
    f_{k|k^\prime}^p\left(\mathbf{x}_k^u\right) &= \exp\left(-\left\langle \lambda_{k|k^\prime},1\right\rangle\right)\left[ \lambda_{k|k^\prime}(\cdot) \right]^{\mathbf{x}_k^u},\label{eq_ppp}\\
    f_{k|k^\prime}^{mbm}\left(\mathbf{x}_k^d\right) &=  \sum_{a\in\mathcal{A}_{k|k^\prime}}w^a_{k|k^\prime}\sum_{\uplus_{l=1}^{n_{k|k^\prime}}\mathbf{x}_k^l = \mathbf{x}_k^d}\prod_{i=1}^{n_{k|k^\prime}}f^{i,a^i}_{k|k^\prime}\left(\mathbf{x}_k^i\right),\label{eq_mbm}
\end{align}
where the PPP $f_{k|k^\prime}^p(\cdot)$ has intensity $\lambda_{k|k^\prime}(\cdot)$, and it represents undetected objects that are hypothesized to exist but have not been detected, whereas the MBM $f_{k|k^\prime}^{mbm}(\cdot)$ represents potential objects that have been detected at least once at some point up to time step $k^\prime$.

The MBM in \eqref{eq_mbm} has $n_{k|k^\prime}$ Bernoulli components. For the $i$-th Bernoulli component, there are $h^i_{k|k^\prime}$ local hypotheses, and its density under local hypothesis $a^i\in\{1,\dots,h^i_{k|k^\prime}\}$ is 
\begin{equation}
    f^{i,a^i}_{k|k^\prime}\left(\mathbf{x}_k^i\right) = \begin{cases}
        1 - r_{k|k^\prime}^{i,a^i} & \mathbf{x}_k^i = \emptyset \\
        r_{k|k^\prime}^{i,a^i} f^{i,a^i}_{k|k^\prime}(x) & \mathbf{x}_k^i = \{x\} \\
        0 & \text{otherwise},
    \end{cases}
\end{equation}
where $r^{i,a^i}_{k|k^\prime}$ is the probability of existence and $f^{i,a^i}_{k|k^\prime}(\cdot)$ is the single object density conditioned on existence. The set $\mathcal{A}_{k|k^\prime}$ contains all the global hypotheses \cite{pmbmextended2}. Each MB component in \eqref{eq_mbm} represents the density of the set $\mathbf{x}_k^d$ of detected objects conditioned on a global hypothesis $a =(a^1,\dots,a^{n_{k|k^\prime}}) \in \mathcal{A}_{k|k^\prime}$, which selects a local hypothesis from each Bernoulli component. The weight of global hypothesis $a$ satisfies 
\begin{equation}
    w^a_{k|k^\prime} \propto \prod_{i=1}^{n_{k|k^\prime}}w^{i,a^i}_{k|k^\prime},
\end{equation}
where $w^{i,a^i}_{k|k^\prime}$ is the weight of local hypothesis $a^i$ for the $i$-th Bernoulli component. Normalization is required to ensure that $\sum_{a\in\mathcal{A}_{k|k^\prime}}w^a_{k|k^\prime} = 1$.

\subsection{Poisson Multi-Bernoulli Approximation}

The PMB is a special case of PMBM with a single MB. By augmenting the object space with auxiliary variables, the best PMB approximation can be obtained by minimizing the Kullback-Leibler divergence (KLD) \cite{garcia2020trajectory}. To do so, we first augment the single object space with an auxiliary variable $u\in\mathbb{U}_{k|k^\prime}=\{0,1,\dots,n_{k|k^\prime}\}$, such that $(u, x)\in \mathbb{U}_{k|k^\prime}\times \mathbb{R}^{d_x}$. The auxiliary variable $u$ can be interpreted as a track index: for $u=i\in\{1,\dots,n_{k|k^\prime}\}$, $x$ represents the state of the $i$-th potential object (also referred to as track in \cite{pmbmpoint}), and for $u=0$, $x$ represents an undetected object. A set of objects equipped with track indices is then denoted as $\widetilde{\mathbf{x}}_k\in\mathcal{F}(\mathbb{U}_{k|k^\prime}\times\mathbb{R}^{d_x})$. Given $f_{k|k^\prime}(\cdot)$ of the form \eqref{eq_pmbm}, the density ${f}_{k|k^\prime}(\cdot)$ on the space ${\cal F}(\mathbb{U}_{k|k^\prime}\times \mathbb{R}^{d_x})$ is \cite{garcia2020trajectory}
\begin{equation}\label{eq_pmbm_au}
  {f}_{k|k^\prime}\left(\widetilde{{\bf x}}_k\right) = {f}^p_{k|k^\prime}\left(\widetilde{{\bf y}}_k\right)\sum_{a\in{\cal A}_{k|k^\prime}}w^a_{k|k^\prime}\prod_{i=1}^{n_{k|k^\prime}}{f}^{i,a^i}_{k|k^\prime}\left(\widetilde{{\bf x}}^i_k\right)
\end{equation}
where, for a given $\widetilde{{\bf x}}_k$, $\widetilde{{\bf y}}_k = \{(u,x)\in \widetilde{{\bf x}}_k: u =0\}$, $\widetilde{{\bf x}}^i_k = \{(u,x)\in \widetilde{{\bf x}}_k: u =i\}$, and 
\begin{subequations}\label{eq_pmb_approx}
\begin{align}
  {f}^p_{k|k^\prime}\left(\widetilde{{\bf y}}_{k}\right) &= \exp\left(-\left\langle {\lambda}_{k|k^\prime},1 \right\rangle\right) \left[ \lambda_{k|k^\prime}(\cdot) \right]^{\widetilde{{\bf y}}_{k}},\\
  {\lambda}_{k|k^\prime}(u,x) &= \delta_0[u]{\lambda}_{k|k^\prime}(x),\\
  {f}^{i,a^i}_{k|k^\prime}\left(\widetilde{{\bf x}}_k^i\right) &= \begin{cases}
    1 - r^{i,a^i}_{k|k^\prime} & \widetilde{{\bf x}}_k^i = \emptyset\\
    \delta_i[u]r^{i,a^i}_{k|k^\prime}f^{i,a^i}_{k|k^\prime}(x) & \widetilde{{\bf x}}_k^i = \{(u,x)\}\\
    0 & \text{otherwise}.
  \end{cases}
\end{align}
\end{subequations}
We notice that the sum over sets in \eqref{eq_pmbm} disappears in \eqref{eq_pmbm_au} due to the use of auxiliary variables since now there is only one possible partition of $\widetilde{{\bf x}}_k$ into $\widetilde{{\bf y}}_k$, $\widetilde{{\bf x}}^1_k,\dots,\widetilde{{\bf x}}^{n_{k|k^\prime}}_k$ that provides a non-zero density. If we integrate out the auxiliary variable $u$ for each object, we recover the original PMBM density \eqref{eq_pmbm} \cite{garcia2020trajectory}. We also note that in $\widetilde{\mathbf{x}}_k$ there can be multiple objects with $u=0$, but at most one object with $u=i\in\{1,\dots,n_{k|k^\prime}\}$, for a non-zero density of \eqref{eq_pmbm_au}.

Given a PMBM density ${f}_{k|k^\prime}(\cdot)$ of the form \eqref{eq_pmbm_au}, the PMB density that minimizes the KLD from ${f}_{k|k^\prime}(\cdot)$ has closed form and is given by \cite[Prop. 2]{garcia2020trajectory}
\begin{subequations}\label{eq_pmb}
  \begin{align}
    {f}^{pmb}_{k|k^\prime}\left(\widetilde{{\bf x}}_{k}\right) &=  {f}^p_{k|k^\prime}\left(\widetilde{{\bf y}}_k\right)\prod_{i=1}^{n_{k|k^\prime}}{f}^{i}_{k|k^\prime}\left(\widetilde{{\bf x}}^i_k\right),\label{eq_pmb_predict}\\
    {f}^{i}_{k|k^\prime}\left(\widetilde{{\bf x}}\right) &= \begin{cases}
      1 - r^{i}_{k|k^\prime} & \widetilde{{\bf x}} = \emptyset\\
      \delta_i[u]r^{i}_{k|k^\prime}f^{i}_{k|k^\prime}(x) & \widetilde{{\bf x}} = \{(u,x)\}\\
      0 & \text{otherwise},
    \end{cases}\label{eq_marginal_ber}\\
    r^i_{k|k^\prime} &=  \sum_{a^i=1}^{h^i_{k|k^\prime}}\overline{w}_{k|k^\prime}^{i,a^i}r^{i,a^i}_{k|k^\prime},\label{eq_pmb_r}\\
    f^i_{k|k^\prime}(x) &=   \sum_{a^i=1}^{h^i_{k|k^\prime}}\overline{w}_{k|k^\prime}^{i,a^i}r^{i,a^i}_{k|k^\prime}f_{k|k^\prime}^{i,a^i}(x)/r^i_{k|k^\prime},\label{eq_pmb_f}\\
    \overline{w}_{k|k^\prime}^{i,a^i} &=  \sum_{b\in {\cal A}_{k|k^\prime}:b^i=a^i}w^b_{k|k^\prime}\label{eq_marginal_assoc}. 
  \end{align}
\end{subequations}
If we integrate out the auxiliary variables in \eqref{eq_pmb}, we obtain the PMB density without auxiliary variables \cite{garcia2020trajectory}
\begin{equation}\label{eq_pmb_original}
  {f}^{pmb}_{k|k^\prime}({{\bf x}}_{k}) = \sum_{\uplus_{l=1}^{n_{k|k^\prime}}{\bf x}^l\uplus {\bf y} = {\bf x}_k} {f}^p_{k|k^\prime}({{\bf y}})\prod_{i=1}^{n_{k|k^\prime}}{f}^{i}_{k|k^\prime}\left({{\bf x}}^i\right)
\end{equation}
where the $i$-th Bernoulli component $f^i_{k|k^\prime}({\bf x}^i)$ is parameterized by $r^i_{k|k^\prime}$ in \eqref{eq_pmb_r} and $f^i_{k|k^\prime}(x)$ in \eqref{eq_pmb_f}.

\subsection{Poisson Multi-Bernoulli Filtering}

We first define the set of global hypotheses for extended objects. We refer to measurement ${\bf z}_k^j$ using the pair $(k,j)$, and the set of all such pairs at time step $k$ is denoted ${\cal M}_k$. We also let ${\cal M}_k^{i,a^i}\subseteq {\cal M}_k$ be the set of index pairs associated to local hypothesis $a^i$ of the $i$-th Bernoulli component. Then, the set of all feasible global hypotheses is 
\begin{equation}
    {\cal A}_{k|k} = 
    \left\{ (a^1,\dots,a^{n_{k|k}}):a^i\in \{1,\dots,h^i_{k|k}\}~\forall~i,\biguplus_{i=1}^{n_{k|k}}{\cal M}_k^{i,a^i} = {\cal M}_k\right\},\label{eq_globalassocconstraint}
\end{equation}
where the constructions of $\mathcal{M}_k^{i,a^i}$ will be given in Lemma~\ref{pmbm_update}. We consider the PMB update, where each measurement creates a new Bernoulli component \cite{xia2021poisson}, instead of each non-empty subset of measurements creating a new Bernoulli component, as in \cite{pmbmextended2}.

\begin{lemma}\label{pmbm_prediction}
    Given the PMB filtering density at time $k-1$ of the form \eqref{eq_pmb_original}, the predicted density at time $k$ is a PMB of the form \eqref{eq_pmb_original}, with $n_{k|k} = n_{k|k-1}$,
    \begin{subequations}
        \begin{align}
            \lambda_{k|k-1}(x) &= \lambda^B_k(x) + \left\langle \lambda_{k-1|k-1},g_k(x|\cdot)p^S(\cdot) \right\rangle,\label{eq_ppp_predict}\\
            r^{i}_{k|k-1} &= r^{i}_{k-1|k-1}\left\langle f^{i}_{k-1|k-1},p^S \right\rangle,\\
            f^{i}_{k-1|k-1}(x) &= \frac{\left\langle f^{i}_{k-1|k-1},g_k(x|\cdot)p^S(\cdot) \right\rangle}{\left\langle f^{i}_{k-1|k-1},p^S \right\rangle}.
        \end{align}
    \end{subequations}
\end{lemma}
\begin{lemma}\label{pmbm_update}
    Given the PMB predicted density at time step $k$ of the form \eqref{eq_pmb_original} and measurements ${\bf z}_k = \{z_k^1,\dots,z_k^{m_k}\}$, the updated density is a PMBM of the form \eqref{eq_pmbm}, with
\begin{align}
  n_{k|k} &= n_{k|k-1} + m_k, \\
  \mathcal{M}_k &= \left\{(k,j) | j \in \{1,\dots,m_k\} \right\},\\
  \lambda_{k|k}(x) &= \ell_k(\emptyset|x)\lambda_{k|k-1}(x).\label{eq_likelihood_missed_ppp}
\end{align}

The $i$-th existing Bernoulli component, $i\in \{1,\dots,n_{k|k-1}\}$, has $h^i_{k|k} = 2^{m_k}$ local hypotheses, either corresponding to a misdetection or an update using a non-empty subset of ${\bf z}_k$. For a misdetection hypothesis, we have $\mathcal{M}_k^{i,1} = \emptyset$, and
\begin{subequations}
  \begin{align}
    \ell_{k|k}^{i,1,0} &= \left\langle f_{k|k-1}^{i},\ell_k(\emptyset|\cdot) \right\rangle,\label{eq_likelihood_missed}\\
    w_{k|k}^{i,1} &= 1 - r_{k|k-1}^{i} + r_{k|k-1}^{i}\ell_{k|k}^{i,1,0},\\
    r_{k|k}^{i,1} &= \frac{r_{k|k-1}^{i}\ell_{k|k}^{i,1,0}}{1 - r_{k|k-1}^{i} + r_{k|k-1}^{i}\ell_{k|k}^{i,1,0}},\\
    f_{k|k}^{i,1}(x) &= \ell_k(\emptyset|x)f_{k|k-1}^{i}(x)/\ell_{k|k}^{i,1,0}.
  \end{align}
\end{subequations}

Let ${\bf w}_k^1,\dots,{\bf w}_k^{2^{m_k}-1}$ be the non-empty subsets of ${\bf z}_k$. For the $i$-th Bernoulli component $i\in\{1,\dots,n_{k|k-1}\}$, the local hypothesis generated by a set ${\bf w}_k^j$ of measurements has index $a^i = 1 + j$ with $j\in\{1,\dots,2^{m_k}-1\}$, and
\begin{subequations}
  \begin{align}
    \mathcal{M}_{k}^{i,a^i} &= \left\{ (k,p): z_k^p \in {\bf w}_k^j \right\},\\
    \ell_{k|k}^{i,a^i,j} &= \left\langle f_{k|k-1}^{i},\ell_k\left({\bf w}_k^j|\cdot\right) \right\rangle, \label{eq_likelihood_Bernoulli_update}\\
    w_{k|k}^{i,a^i} &= r_{k|k-1}^{i}\ell_{k|k}^{i,a^i,j},\\
    r_{k|k}^{i,a^i} &= 1,\\
    f_{k|k}^{i,a^i}(x) &= \ell_k\left({\bf w}_k^j|x\right)f_{k|k-1}^{i}(x)/\ell_{k|k}^{i,a^i,j}.
  \end{align}
\end{subequations}

Each new Bernoulli component has a different number of local hypotheses, each created by a subset of ${\bf z}_k$.  The set of subsets of measurements associated to the $i$-th new Bernoulli component ($i\in \{1,\dots,m_k\}$) can be recursively built as \cite{xia2021poisson}
\begin{equation}\label{eq_new_local_meas}
  {\cal S}_i = \left\{ \left\{ {z}_k^i \right\} \right\} \cup \left( \cup_{{\bf w}\in \cup_{j=1}^{i-1}{\cal S}_j} \left\{ \left\{ { z}_k^i \right\} \cup {\bf w}  \right\} \right),
\end{equation}
with ${\cal S}_1 = \{ \{ {z}_k^1 \} \}$. This implies that the $i$-th new Bernoulli component must contain the local hypothesis created by the single measurement $z_k^i$ and cannot be associated with measurements with index larger than $i$. Further, we let ${\bf w}_k^{i,\iota} \in \mathcal{S}_i$ denote the $\iota$-th non-empty subset of measurements associated to the $i$-th new Bernoulli component, with $i = n_{k|k-1}+j$, $j\in\{1,\dots,m_k\},\iota\in \{1,\dots,2^{j-1}\}$. The $i$-th new Bernoulli component has $h^i_{k|k} = 2^{j-1}+1$ local hypotheses. The case corresponding to  non-existence is given by
\begin{equation}\label{eq_non_existence_ber}
  {\cal M}_k^{i,1} = \emptyset, \quad w_{k|k}^{i,1} = 1, \quad r_{k|k}^{i,1} = 0,
\end{equation}
and the others ($a^i = \iota+1,\iota\in \{1,\dots,2^{j-1}\}$), corresponding to detection with ${\bf w}_k^{i,\iota}$, are given by
\begin{subequations}\label{eq_newBernoulli2}
  \begin{align}
    {\cal M}_k^{i,a^i} &= \left\{ (k,p):z_k^p\in {\bf w}_k^{i,\iota} \right\},\\
    \ell_{k|k}^{i,\iota} &= \left\langle \lambda_{k|k-1},\ell_k\left({\bf w}_k^{i,\iota}|\cdot\right) \right\rangle,\label{eq_detection_likelihood_undetected}\\
    w_{k|k}^{i,a^i} &= \delta_1\left[\left|{\bf w}_k^{i,\iota}\right|\right]\lambda_k^C\left(z_k^j\right)+ \ell_{k|k}^{i,\iota},\\
    r_{k|k}^{i,a^i} &= \ell_{k|k}^{i,\iota}/w_{k|k}^{i,a^i},\\
    f_{k|k}^{i,a^i}(x) &= \ell_k\left({\bf w}_k^{i,\iota}|x\right)\lambda_{k|k-1}(x)/\ell_{k|k}^{i,\iota}.
  \end{align}
\end{subequations}
\end{lemma}

In \eqref{eq_newBernoulli2}, under the local hypothesis with $\mathbf{w}_k^{i,\iota} = \{z_k^j\}$, the Bernoulli set density accounts for the possibility that the measurement $z_k^j$ is either clutter or generated by a newly detected object. If $|{\bf w}_k^{i,\iota}|>1$, the local hypothesis corresponds to an object with existence probability one.

\section{Joint Posterior and its Factorization}\label{sec_joint_posterior}

In this section, we first present the joint posterior of the set of objects and global hypothesis. Then, we describe how to factorize this joint posterior for the ZIP measurement model, making it amenable to message passing.

\subsection{Joint Posterior of Set of Objects and Global Hypothesis}

We define the joint posterior of the set of objects $\widetilde{\mathbf{x}}_k$ and global hypothesis $a$ on the space ${\cal F}(\mathbb{U}_{k|k}\times \mathbb{R}^{d_x}) \times \mathcal{A}_{k|k}$ as 
\begin{equation}
    {f}_{k|k}\left(\widetilde{\mathbf{x}}_k,a\right) =  {f}_{k|k}^p\left(\widetilde{\mathbf{x}}_k^u\right) w^a_{k|k} \prod_{i=1}^{n_{k|k}}{f}^{i,a^i}_{k|k}\left(\widetilde{\mathbf{x}}_k^i\right)\propto  {f}_{k|k}^p\left(\widetilde{\mathbf{x}}_k^u\right) \prod_{i=1}^{n_{k|k}} w^{i,a^i}_{k|k} {f}^{i,a^i}_{k|k}\left(\widetilde{\mathbf{x}}_k^i\right)\triangleq  {f}_{k|k}^p\left(\widetilde{\mathbf{x}}_k^u\right) \prod_{i=1}^{n_{k|k}} {g}^{i,a^i}_{k|k}\left(\widetilde{\mathbf{x}}_k^i\right),
    \label{eq_joint_posterior}
\end{equation}
where ${g}^{i,a^i}_{k|k}(\cdot)$ correspond to unnormalized Bernoulli densities and can be computed as follows according to Lemma~\ref{pmbm_update}. Specifically, we have that, for missed detection hypotheses, with $i\in \{1,\dots,n_{k|k-1}\}$, 
\begin{equation}
  {g}^{i,1}_{k|k}\left(\widetilde{\mathbf{x}}^i_k\right) = \begin{cases}
    \delta_i[u]r^{i}_{k|k-1}\ell_k(\emptyset|x)f^{i}_{k|k-1}(x) & \widetilde{\mathbf{x}}^i_k = \{(u,x)\} \\
    1-r^{i}_{k|k-1} & \widetilde{\mathbf{x}}^i_k = \emptyset \\
    0 & \text{otherwise},
  \end{cases}
\end{equation}
and for detection hypotheses, with $a^i = 1 + j$, where $j\in\{1,\dots,2^{m_k}-1\}$,
\begin{equation}
  {g}^{i,a^i}_{k|k}\left(\widetilde{\mathbf{x}}^i_k\right) =\begin{cases}
     \delta_i[u]r^{i}_{k|k-1}\ell_k({\bf w}_k^j|x)f_{k|k-1}^{i}(x) & \widetilde{\mathbf{x}}_k^i = \{(u,x)\} \\
     0 & \text{otherwise}.
  \end{cases}
\end{equation}
For the non-existence hypotheses for new Bernoulli components, with $i = n_{k|k-1}+j,j\in\{1,\dots,m_k\}$,
\begin{equation}
  {g}^{i,1}_{k|k}\left(\widetilde{\mathbf{x}}^i_k\right) = \begin{cases}
    1 & \widetilde{\mathbf{x}}^i_k = \emptyset \\ 
    0 & \text{otherwise},
  \end{cases}
\end{equation}
and for the existence hypotheses for new Bernoulli components, with $a^i = \iota+1$, $\iota\in \{1,\dots,2^{j-1}\}$
\begin{equation}
  {g}^{i,a^i}_{k|k}\left(\widetilde{\mathbf{x}}^i_k\right) 
  = \begin{cases}
    \delta_i[u]\ell_k\left({\bf w}_k^{i,\iota}|x\right)\lambda_{k|k-1}(x) & \widetilde{\mathbf{x}}^i_k = \{(u,x)\} \\ 
   \delta_1\left[\left|{\bf w}_k^{i,\iota}\right|\right]\lambda_k^C\left(z_k^j\right) & \widetilde{\mathbf{x}}^i_k = \emptyset \\
    0 & \text{otherwise}.
  \end{cases}
\end{equation}

It can be observed that the joint posterior of the set of objects and the global hypothesis in \eqref{eq_joint_posterior} already admits a factorized form. Nevertheless, this form is not directly amenable to BP, since the global hypothesis $a\in\mathcal{A}_{k|k}$ is subject to the constraint in \eqref{eq_globalassocconstraint}, which acts as an implicit high-order factor that couples the local hypotheses globally, thereby preventing a convenient decomposition into low-order factors for efficient BP.

\subsection{Joint Posterior Factorization with Auxiliary Variables}

To make the posterior more amenable to message passing, we first introduce binary object detection variables and dual representations of object-oriented and measurement-oriented data association variables in order to expose the latent structure of the PMB update with ZIP measurements. Specifically, for the $i$-th existing Bernoulli component, $i\in\{1,\dots,n_{k|k-1}\}$, we first introduce a binary object detection variable $D^i_k\in\{0,1\}$, where $D_k^i=1$ indicates that the object is detected at time step $k$ (and generates a Poisson number $\gamma_k(\cdot)$ of measurements which can still be zero) while $D^i_k=0$ indicates a missed detection that corresponds to the Bernoulli non-detection branch. Note that we do not need to introduce $D_k^i$ for new Bernoulli components that represent newly detected objects. Then, we have
\begin{equation}
  \label{eq_condition_xD}
  \ell_k^D\left(\mathbf{w}_k,D_k^i | \widetilde{\mathbf{x}}_k^i\right) = \ell_k\left(\mathbf w_k | \widetilde{\mathbf{x}}_k^i, D_k^i\right)f_k^{D,i}\left(D_k^i | \widetilde{\mathbf{x}}_k^i\right),
\end{equation}
where 
\begin{align}
  \label{eq_detection_def}
  f_k^{D,i}\left(D_k^i | \widetilde{\mathbf{x}}_k^i\right) &= \begin{cases}
    \delta_i[u]p^D_k(x)e^{-\gamma_k(x)} & \widetilde{\mathbf{x}}_k^i = \{(u,x)\}, D_k^i = 1 \\
   \delta_i[u] \left(1 - p^D_k(x) \right) & \widetilde{\mathbf{x}}_k^i = \{(u,x)\}, D_k^i = 0 \\
    1 & \widetilde{\mathbf{x}}_k^i = \emptyset, D_k^i = 0 \\
    0 & \text{otherwise},
  \end{cases}\\
\ell_k\left(\mathbf w_k | \widetilde{\mathbf{x}}_k^i = \{(u,x)\}, D_k^i\right) 
&=
\begin{cases}
 \delta_i[u] \left[\gamma_k(x)\ell_k(\cdot|x)\right]^{\mathbf{w}_k} & D_k^i=1 \\
\delta_i[u] & D_k^i = 0, \mathbf w_k=\emptyset\\
0 & \text{otherwise},
\end{cases}\\
\ell_k\left(\mathbf w_k | \widetilde{\mathbf{x}}_k^i = \emptyset, D_k^i\right) &=
\begin{cases}
1 & D_k^i = 0, \mathbf w_k=\emptyset\\
0 & \text{otherwise}. 
\end{cases}
\end{align}

If we marginalize $D_k^i$ from \eqref{eq_condition_xD}, we recover the original ZIP measurement set density. By separating the Bernoulli detection event from the conditional Poisson measurements, we obtain a more explicit factorized representation \eqref{eq_condition_xD}.

We proceed to describe how to decompose the high-order factor introduced by the global association hypothesis constraint $a\in\mathcal{A}_{k|k}$ into a collection of lower-order factors suitable for BP. To do so, we introduce a dual representation of the data association, one object-oriented and one measurement-oriented. Although redundant, both representations are equivalent, in the sense that each admissible realization of either representation under the imposed constraints uniquely determines a global hypothesis \cite[App. B]{xia2023trajectory}. This redundancy allows the original global hypothesis constraint in \eqref{eq_globalassocconstraint} to be replaced by local consistency factors.

We first introduce the object-oriented association variables
$\alpha_k=[(\alpha_k^1)^T,\dots,(\alpha_k^{n_{k|k}})^T]^T$,
where
\begin{equation}
  \label{eq_object_oriented_variable}
\alpha_k^i=
\begin{cases}
\left[\alpha_k^{i,1},\dots,\alpha_k^{i,m_k}\right]^T
& i\in\{1,\dots,n_{k|k-1}\}\\[2mm]
\left[\alpha_k^{i,1},\dots,\alpha_k^{i,j}\right]^T
& i = n_{k|k-1}+j,
j \in\{1,\dots,m_k\},
\end{cases}
\end{equation}
with $\alpha_k^{i,j}\in\{0,1\}$, and $\alpha_k^{i,j} = 1$ if and only if the $j$-th measurement $z_k^j$ is associated with the $i$-th Bernoulli component. Note that the object-oriented association vector corresponding to the $j$-th new Bernoulli component has length $j$. This follows directly from the local hypothesis structure of new Bernoulli components described in Lemma~\ref{pmbm_update}.

In addition, according to the conditional measurement set density \eqref{eq_condition_xD}, if the potential object corresponding to the $i$-th Bernoulli component is not detected, then no measurement can be associated with it. To encode this relation, we introduce the binary factor
\begin{equation}
  \label{binary_factor_detection}
  \Phi_k^{i,j}\left(D_k^i,\alpha_k^{i,j}\right) =
\begin{cases}
1 & D_k^i=1\ \text{or}\ \alpha_k^{i,j}=0\\
0 & \text{otherwise},
\end{cases}
\end{equation}
for $i\in\{1,\dots,n_{k|k-1}\}$ and $j\in\{1,\dots,m_k\}$.  The product of these factors over index $j$ ensures that no measurement can be associated with the $i$-th Bernoulli component when $D_k^i=0$, while the case with $D_k^i=1$ and all-zero $\alpha_k^{i,j}$ is still allowed, which represents that the corresponding object is detected but generates zero measurements.

We proceed to rewrite the unnormalized Bernoulli densities
${g}^{i,a^i}_{k|k}(\cdot)$ using the object-oriented association variables $\alpha_k$ and binary detection variables $D_k^{1:n_{k|k-1}} = [D_k^1,\dots,D_k^{n_{k|k-1}}]^T$, leveraging the factorization of the conditional Poisson measurement set density \eqref{eq_condition_xD}. Specifically, for the $i$-th Bernoulli component with $i\in\{1,\dots,n_{k|k-1}\}$, we have 
\begin{equation}
  \label{eq_g_exist}
  {g}^{D,i,\alpha_k^i}_{k|k}\left(\widetilde{\mathbf{x}}_k^i,D_k^i\right) = f_k^{D,i}\left(D_k^i | \widetilde{\mathbf{x}}_k^i\right)\underline{f}^i_{k|k-1}\left(\widetilde{\mathbf{x}}_k^i\right) \prod_{j=1}^{m_k}\Phi_k^{i,j}\left(D_k^i,\alpha_k^{i,j}\right)\underline{s}_k^i\left(\widetilde{\mathbf{x}}_k^i,\alpha_k^{i,j};z_k^j\right),
\end{equation}
and for $i = n_{k|k-1}+j$, $j\in\{1,\dots,m_k\}$,
\begin{equation}
  \label{eq_g_new}
  {g}^{i,\alpha_k^i}_{k|k}\left(\widetilde{\mathbf{x}}_k^i\right) = \overline{f}^i_{k|k-1}\left(\widetilde{\mathbf{x}}_k^i\right) \overline{s}_k^i\left(\widetilde{\mathbf{x}}_k^i,\alpha_k^{i,j};z_k^j\right)\prod_{l=1}^{j-1}\underline{s}_k^i\left(\widetilde{\mathbf{x}}_k^i,\alpha_k^{i,l};z_k^l\right).
\end{equation}
In \eqref{eq_g_exist} and \eqref{eq_g_new}, the local hypothesis $a^i$ of the $i$-th Bernoulli component is now represented explicitly by the corresponding object-oriented association variable $\alpha_k^i$ \eqref{eq_object_oriented_variable}, which provides an equivalent description of the measurement subset associated with that component. In addition, the new factors are
\begin{align}
  \label{eq_prior_exist}
  \underline{f}^i_{k|k-1}\left(\widetilde{\mathbf{x}}_k^i\right) &= \begin{cases}
    \delta_i[u]r^i_{k|k-1}f^i_{k|k-1}(x)& \widetilde{\mathbf{x}}_k^i = \{(u,x)\} \\
    1 - r^i_{k|k-1} & \widetilde{\mathbf{x}}_k^i = \emptyset \\
    0 & \text{otherwise},
  \end{cases}\\
  \label{eq_prior_new}
  \overline{f}^i_{k|k-1}\left(\widetilde{\mathbf{x}}_k^i\right) 
  &= \begin{cases}
    \delta_i[u]p^D_k(x)e^{-\gamma_k(x)}\lambda_{k|k-1}(x) & \widetilde{\mathbf{x}}_k^i = \{(u,x)\} \\
    1 & \widetilde{\mathbf{x}}_k^i = \emptyset \\
    0 & \text{otherwise},
  \end{cases}\\
  \label{eq_meas_exist}
  \underline{s}_k^i\left(\widetilde{\mathbf{x}}_k^i,\alpha_k^{i,j};z_k^j\right) 
  &= \begin{cases}
    \delta_i[u]\gamma_k(x)\ell_k\left(z_k^j | x\right) & \widetilde{\mathbf{x}}_k^i = \{(u,x)\}, \alpha_k^{i,j} = 1 \\
    1 & \alpha_k^{i,j} = 0 \\
    0 & \text{otherwise},
  \end{cases}\\
  \label{eq_meas_new}
  \overline{s}_k^i\left(\widetilde{\mathbf{x}}_k^i,\alpha_k^{i,j};z_k^j\right) 
  &= \begin{cases}
    \delta_i[u]\gamma_k(x)\ell_k\left(z_k^j | x\right) & \widetilde{\mathbf{x}}_k^i = \{(u,x)\}, \alpha_k^{i,j} = 1 \\
    \lambda_k^C\left(z_k^j\right) & \widetilde{\mathbf{x}}_k^i = \emptyset, \alpha_k^{i,j} = 1 \\
    1 & \widetilde{\mathbf{x}}_k^i = \emptyset, \alpha_k^{i,j} = 0  \\
    0 & \text{otherwise}.
  \end{cases}
\end{align}
In the above, the factors \eqref{eq_prior_exist} and \eqref{eq_prior_new} represent the predicted and prior information for existing and newly detected objects, respectively, without being associated with any measurements, whereas the factors \eqref{eq_meas_exist} and \eqref{eq_meas_new} represent the local likelihood of associating a particular measurement with its
corresponding Bernoulli component for existing and newly detected objects, respectively.

Next, we introduce the measurement-oriented association variables $\beta_k = [\beta_k^1,\dots,\beta_k^{m_k}]^T$, where the $j$-th element $\beta_k^j=i$, $j\in\{1,\dots,m_k\}$, if and only if the $j$-th measurement $z_k^j$ is associated with the $i$-th Bernoulli component. According to the local hypothesis structure for new Bernoulli components in Lemma~\ref{pmbm_update}, the $j$-th measurement cannot be associated with a new Bernoulli component, whose index is smaller than $n_{k|k-1}+j$. Thus, for each $j\in\{1,\dots,m_k\}$,
$\beta_k^j \in \{1,\dots,n_{k|k-1}, n_{k|k-1}+j,\dots,n_{k|k}\}$.

To eliminate the implicit high-order factor induced by the global association hypothesis constraint, we introduce low-order consistency factors between the object-oriented $\alpha_k$ and the measurement-oriented $\beta_k$ association variables. Specifically, for each valid pair $(i,j)$, we define
\begin{equation}
  \label{eq_constraint_factor}
  \Psi_k^{i,j}\left(\alpha_k^{i,j},\beta_k^j\right) =
\begin{cases}
1 & \alpha_k^{i,j}=1, \beta_k^j = i \\ 
1 & \alpha_k^{i,j}=0, \beta_k^j\neq i\\
0 & \text{otherwise}.
\end{cases}
\end{equation}
This factor enforces that the dual representations of the data association agree locally on whether the $j$-th measurement $z_k^j$ is associated with the $i$-th Bernoulli component. Consequently, the original global hypothesis constraint in \eqref{eq_globalassocconstraint}
is now represented explicitly through a product of consistency factors between the dual association variables, thereby transforming the remaining high-order dependency into a BP-friendly local factorization.

\begin{figure}[!t]
    \centering
    \includegraphics[width=0.95\columnwidth]{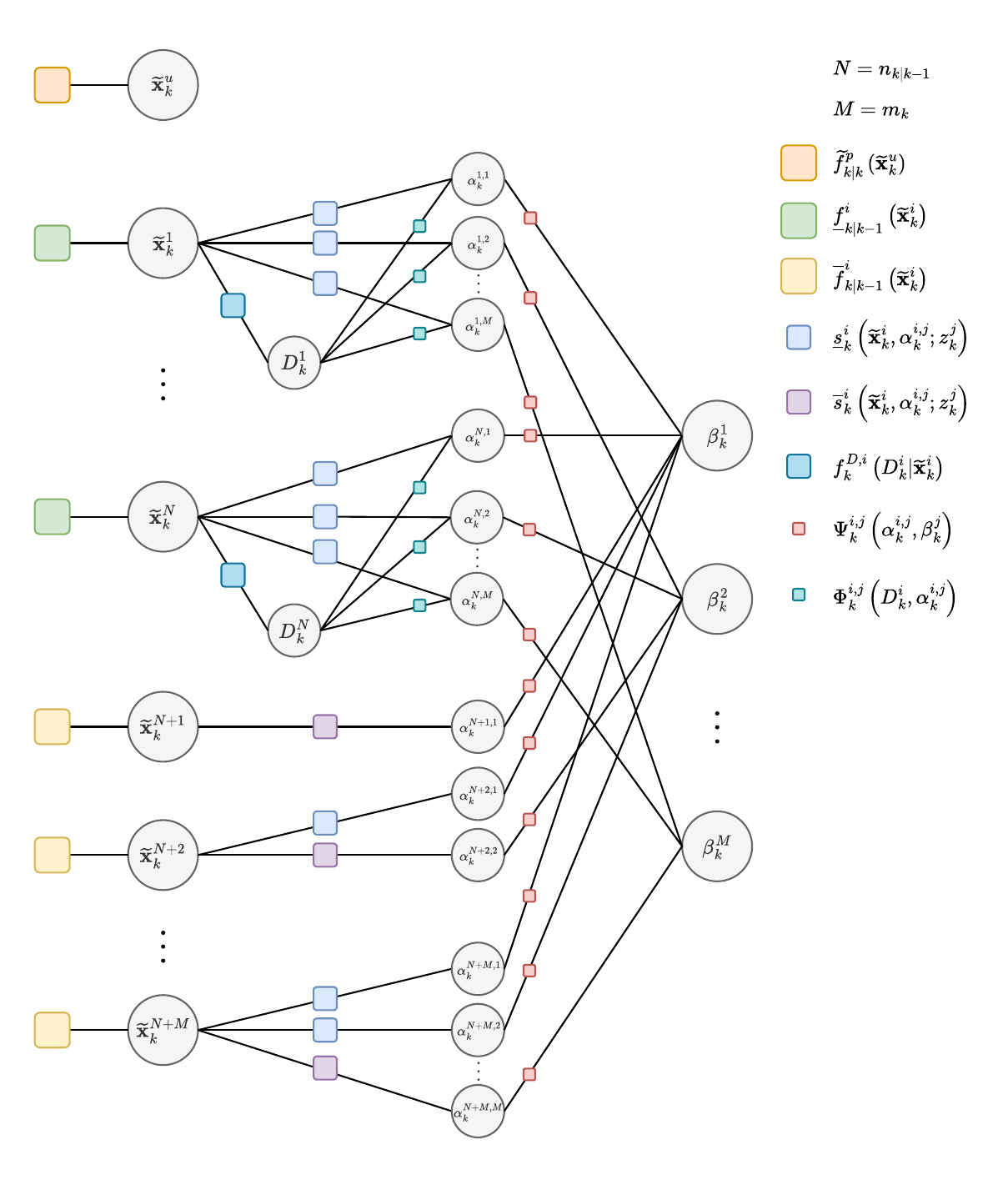}
    \caption{Factor graph for the factorized joint posterior \eqref{eq_full_factorize}, where variable and factor nodes are represented using circles and squares, respectively.}
    \label{fig_factor_graph}
\end{figure}

By plugging \eqref{eq_g_exist} and \eqref{eq_g_new} into \eqref{eq_joint_posterior} and using the low-order consistency factors between the dual representation of data association, we can factorize the joint posterior of the set of objects $\widetilde{\mathbf{x}}_k$, object detection variables $D_k^{1:n_{k|k-1}}$, object-oriented $\alpha_k$ and measurement-oriented $\beta_k$ association variables as in \eqref{eq_full_factorize}. The factor graph corresponding to this factorized joint posterior is illustrated in Fig. \ref{fig_factor_graph}.
  \begin{align}
    \label{eq_full_factorize}
  & {f}_{k|k}\left(\widetilde{\mathbf{x}}_k,D_k^{1:n_{k|k-1}},\alpha_k,\beta_k\right)\nonumber \\
  &\propto {f}_{k|k}^p\left(\widetilde{\mathbf{x}}_k^u\right)\prod_{i=1}^{n_{k|k-1}}\left[f_k^{D,i}\left(D_k^i | \widetilde{\mathbf{x}}_k^i\right)\underline{f}^i_{k|k-1}\left(\widetilde{\mathbf{x}}_k^i\right) \prod_{j=1}^{m_k}\Phi_k^{i,j}\left(D_k^i,\alpha_k^{i,j}\right)\right]\prod_{i=n_{k|k-1}+1}^{n_{k|k}}\overline{f}^i_{k|k-1}\left(\widetilde{\mathbf{x}}_k^i\right) \nonumber \\
  &\times  \prod_{j=1}^{m_k}\left[ \overline{s}_k^{n_{k|k-1}+j}\left(\widetilde{\mathbf{x}}_k^{n_{k|k-1}+j},\alpha_k^{n_{k|k-1}+j,j};z_k^{j}\right)\prod_{\substack{i\in\{1,\dots,n_{k|k-1}\}\cup \\ \{n_{k|k-1}+j+1,\dots,n_{k|k}\}}} \underline{s}_k^i\left(\widetilde{\mathbf{x}}_k^i,\alpha_k^{i,j};z_k^j\right) \prod_{\substack{i\in\{1,\dots,n_{k|k-1}\}\cup \\ \{n_{k|k-1}+j,\dots,n_{k|k}\}}} \Psi_k^{i,j}\left(\alpha_k^{i,j},\beta_k^j\right)\right].
\end{align}

\begin{remark}
  In Fig.~\ref{fig_factor_graph}, we can notice that, for each new Bernoulli component $i\in\{n_{k|k-1}+1,\dots,n_{k|k}\}$, each object-oriented association variable node $\alpha_k^{i,j}$ is connected to two factor nodes, $\widetilde{\mathbf{x}}_k^i$ and $\beta_k^j$. This suggests that each variable node $\alpha_k^{i,j}$ and its two neighboring factor nodes can be combined into a single factor node that directly connects variable nodes $\widetilde{\mathbf{x}}_k^i$ and $\beta_k^j$, by marginalizing out the object-oriented association variables $\alpha_k^{i,j}$ for new Bernoulli components from the joint posterior. This observation leads to a simpler but equivalent factor graph for the new Bernoulli components. Nevertheless, for clarity, in the next section we will derive the loopy BP recursions using the factor graph in Fig.~\ref{fig_factor_graph}, which allows the message passing equations for existing and new Bernoulli components to be written in a similar form. In implementation, the relevant messages for new Bernoulli components can still be collapsed by marginalizing out $\alpha_k^{i,j}$, so the computational complexity remains unchanged.
\end{remark}

\begin{remark}
Compared with the factor graph for the Poisson measurement model in \cite[Fig. 2]{xia2023trajectory}, the factor graph for the ZIP model in Fig.~\ref{fig_factor_graph} contains additional ingredients: the detection variables $D_k^{i}$ and factors $f_k^{D,i}$ for the existing Bernoulli components, the object-oriented association variables $\alpha_k$, as well as the consistency factors $\Phi_k^{i,j}$ and $\Psi_k^{i,j}$. The reason is that, for the Poisson measurement model, the object-generated measurement set density factorizes directly over individual measurements, so that the posterior can be expressed solely in terms of the measurement-oriented association variables $\beta_k$. In the ZIP case, however, the object detection event needs to be modeled explicitly, which introduces an additional dependency between object detection and measurement associations. Retaining the  object-oriented association variables therefore provides a convenient way to expose this structure through low-order factors. 
\end{remark}

\section{Loopy Belief Propagation}
\label{sec_lbp}
The objective is to compute the marginal Bernoulli density $f^i_{k|k}(\widetilde{\mathbf{x}}_k^i)$, cf. \eqref{eq_marginal_ber}, for $i\in\{1,\dots,n_{k|k}\}$. Since the factor graph in Fig. \ref{fig_factor_graph} contains cycles, exact BP is no longer applicable, and we instead resort to loopy BP, with all messages updated in parallel as in \cite{florian2021scalable,xia2023trajectory}. In this section, we present the resulting message passing and belief update equations. For notational simplicity, the time index is omitted from all messages. Also, whenever the Kronecker delta function $\delta_i[u]$ has fixed the auxiliary variable to $u=i$, we use the shorthand $\{(i,x)\}$ for the corresponding singleton state.

\subsection{Extrinsic messages from Bernoulli state node to local likelihood factor node}

For an existing Bernoulli component $i\in\{1,\dots,n_{k|k-1}\}$, the extrinsic message from \(\widetilde{\mathbf{x}}_k^i\) to the local likelihood factor \(\underline{s}_k^i(\widetilde{\mathbf{x}}_k^i,\alpha_k^{i,j};z_k^j)\) for $j\in\{1,\dots,m_k\}$ is given by
\begin{equation}
\label{eq_epsilon_exist}
\epsilon^{i,j}\left(\widetilde{\mathbf{x}}_k^i\right)
=
\underline{f}_{k|k-1}^{i}\left(\widetilde{\mathbf{x}}_k^i\right)
\zeta^i\left(\widetilde{\mathbf{x}}_k^i\right)
\prod_{\substack{l=1\\l\neq j}}^{m_k}
\varrho^{i,l}\left(\widetilde{\mathbf{x}}_k^i\right),
\end{equation}
where \(\zeta^i(\widetilde{\mathbf{x}}_k^i)\) represents the message sent from the detection subgraph to \(\widetilde{\mathbf{x}}_k^i\), and \(\varrho^{i,l}(\widetilde{\mathbf{x}}_k^i)\) represents the message from the \(l\)-th local likelihood branch back to \(\widetilde{\mathbf{x}}_k^i\).

For a new Bernoulli component \(i=n_{k|k-1}+j^\prime\), with \(j^\prime\in\{1,\dots,m_k\}\), there is no detection subgraph. In this case, the extrinsic message from \(\widetilde{\mathbf{x}}_k^i\) to the local likelihood factor, \(\underline{s}_k^i\) or \(\overline{s}_k^i\), associated with measurement \(z_k^j\), where \(j\in\{1,\dots,j'\}\), is given by
\begin{equation}
\label{eq_epsilon_new}
\epsilon^{i,j}\left(\widetilde{\mathbf{x}}_k^i\right)
=
\overline{f}_{k|k-1}^{i}\left(\widetilde{\mathbf{x}}_k^i\right)
\prod_{\substack{l=1\\l\neq j}}^{j'}
\varrho^{i,l}\left(\widetilde{\mathbf{x}}_k^i\right).
\end{equation}
The messages \(\epsilon^{i,j}(\widetilde{\mathbf{x}}_k^i)\) provide the state-dependent extrinsic information required to evaluate the compatibility between the \(i\)-th Bernoulli component and the $j$-th measurement \(z_k^j\). 

\subsection{Messages from local likelihood factor node to object-oriented association variable node}

For an existing Bernoulli component \(i\in\{1,\dots,n_{k|k-1}\}\), the message from local likelihood factor \(\underline{s}_k^i(\widetilde{\mathbf{x}}_k^i,\alpha_k^{i,j};z_k^j)\) to \(\alpha_k^{i,j}\) for $j\in\{1,\dots,m_k\}$ is given by
\begin{equation}
\label{eq_theta_exist_def}
\vartheta^{i,j}\left(\alpha_k^{i,j}\right)
=
\int
\underline{s}_k^i\left(\widetilde{\mathbf{x}}_k^i,\alpha_k^{i,j};z_k^j\right)
\epsilon^{i,j}\left(\widetilde{\mathbf{x}}_k^i\right)
\delta \widetilde{\mathbf{x}}_k^i.
\end{equation}
Using \eqref{eq_meas_exist} and normalizing $\vartheta^{i,j}(\alpha_k^{i,j})$ by $\vartheta^{i,j}(0)$, we obtain $\overline{\vartheta}^{i,j}(0) = 1$ and
\begin{equation}
\label{eq_theta_exist}
\overline{\vartheta}^{i,j}(1)
= \frac{\int \gamma_k(x)\ell_k\left(z_k^j | x\right)
\epsilon^{i,j}\left(\{(i,x)\}\right) dx}{\int \epsilon^{i,j}\left(\widetilde{\mathbf{x}}_k^i\right)
\delta \widetilde{\mathbf{x}}_k^i},
\end{equation}
where for a normalized message $\epsilon^{i,j}(\widetilde{\mathbf{x}}_k^i)$, $\int \epsilon^{i,j}(\widetilde{\mathbf{x}}_k^i)
\delta \widetilde{\mathbf{x}}_k^i=1$. Since the BP messages are only defined up to a positive scale factor, the state messages $\epsilon^{i,j}(\widetilde{\mathbf{x}}_k^i)$ are normalized after each update whenever convenient.

For a new Bernoulli component \(i=n_{k|k-1}+j'\), with \(j'\in\{1,\dots,m_k\}\), two cases arise. If \(j<j'\), the corresponding local likelihood factor is \(\underline{s}_k^i(\widetilde{\mathbf{x}}_k^i,\alpha_k^{i,j};z_k^j)\), and the message is also represented by \eqref{eq_theta_exist_def}. If \(j=j'\), the local likelihood factor is \(\overline{s}_k^i(\widetilde{\mathbf{x}}_k^i,\alpha_k^{i,j};z_k^j)\), and we have
\begin{equation}
  \vartheta^{i,j}\left(\alpha_k^{i,j}\right)
=
\int
\overline{s}_k^i \left(\widetilde{\mathbf{x}}_k^i,\alpha_k^{i,j};z_k^j\right)
\epsilon^{i,j} \left(\widetilde{\mathbf{x}}_k^i\right)
\delta \widetilde{\mathbf{x}}_k^i.
\end{equation}
Using \eqref{eq_meas_new} and normalizing $\vartheta^{i,j}(\alpha_k^{i,j})$ by $\vartheta^{i,j}(0)$, we obtain $\overline{\vartheta}^{i,j}(0) = 1$ and
\begin{equation}
\label{eq_theta_new}
\overline{\vartheta}^{i,j}(1)
= \frac{\int \gamma_k(x)\ell_k\left(z_k^j| x\right)
\epsilon^{i,j}\left(\{(i,x)\}\right) dx}{\epsilon^{i,j}(\emptyset)} + \lambda_k^C\left(z_k^j\right).
\end{equation}
The normalized message \(\overline{\vartheta}^{i,j}(\alpha_k^{i,j})\) provides a scalar measure of how compatible the \(i\)-th Bernoulli component is with measurement \(z_k^j\), relative to the inactive case \(\alpha_k^{i,j}=0\). 

\subsection{Messages in the object detection subgraph}

For an existing Bernoulli component $i\in\{1,\dots,n_{k|k-1}\}$, its detection subgraph has the structure
\begin{equation}
\widetilde{\mathbf{x}}_k^i
-
f_k^{D,i}
-
D_k^i
-
\{\Phi_k^{i,j}\}_{j=1}^{m_k}
-
\{\alpha_k^{i,j}\}_{j=1}^{m_k}.
\end{equation}
This subgraph produces two types of messages: one towards the Bernoulli state nodes \(\widetilde{\mathbf{x}}_k^i\), and the other towards the object-oriented association variable nodes \(\alpha_k^{i,j}\). 

For $i\in\{1,\dots,n_{k|k-1}\}$, $j\in\{1,\dots,m_k\}$, we let $\eta_{\Phi}^{i,j}(\alpha_k^{i,j})$
denote the incoming extrinsic message from $\alpha_k^{i,j}$ to the factor \(\Phi_k^{i,j}(D_k^i,\alpha_k^{i,j})\), which is given by
\begin{equation}
\label{eq_eta_phi}
\eta_{\Phi}^{i,j}\left(\alpha_k^{i,j}\right)
= 
\overline{\vartheta}^{i,j}\left(\alpha_k^{i,j}\right)
{\xi}^{i,j}\left(\alpha_k^{i,j}\right) = \begin{cases}
\overline{\vartheta}^{i,j}(1)\xi^{i,j}(1) & \alpha_k^{i,j} = 1 \\
\xi^{i,j}(0) & \alpha_k^{i,j} = 0,
\end{cases}
\end{equation}
where ${\xi}^{i,j}$ is the message from \(\Psi_k^{i,j}\) to \(\alpha_k^{i,j}\), defined later in \eqref{eq_xi_def}.
Then, the message from \(\Phi_k^{i,j}\) to \(D_k^i\) is
\begin{equation}
\label{eq_phi_to_D}
\mu^{i,j}\left(D_k^i\right)
= 
\sum_{\alpha_k^{i,j}\in\{0,1\}}
\Phi_k^{i,j}\left(D_k^i,\alpha_k^{i,j}\right)
\eta^{i,j}_{\Phi}\left(\alpha_k^{i,j}\right) = \begin{cases}
  \xi^{i,j}(0) + \overline{\vartheta}^{i,j}(1)\xi^{i,j}(1) & D_k^i = 1 \\
  \xi^{i,j}(0) & D_k^i = 0.
\end{cases}
\end{equation}
By combining the messages \({\mu}^{i,j}(D_k^i)\) over all \(j\), the message from the detection subgraph to \(\widetilde{\mathbf{x}}_k^i\) is given by
\begin{equation}
\label{eq_zeta_def}
\zeta^i \left(\widetilde{\mathbf{x}}_k^i\right)
= 
\sum_{D_k^i\in\{0,1\}}
f_k^{D,i} \left(D_k^i | \widetilde{\mathbf{x}}_k^i\right)
\prod_{j=1}^{m_k}{\mu}^{i,j}\left(D_k^i\right).
\end{equation}
Plugging \eqref{eq_detection_def} and \eqref{eq_phi_to_D} into \eqref{eq_zeta_def} yields
\begin{equation}
\label{eq_zeta_closed}
\zeta^i\left(\widetilde{\mathbf{x}}_k^i\right) 
=
\begin{cases}
\delta_i[u]\left[
\left(1 - p^D_k(x)\right)\prod_{j=1}^{m_k}{\mu}^{i,j}(0) +   p^D_k(x)e^{-\gamma_k(x)}\prod_{j=1}^{m_k}{\mu}^{i,j}(1)
\right]
& \widetilde{\mathbf{x}}_k^i=\{(u,x)\}\\
\prod_{j=1}^{m_k}{\mu}^{i,j}(0)
& \widetilde{\mathbf{x}}_k^i=\emptyset\\
0
& \text{otherwise}.
\end{cases}
\end{equation}
Hence, $\zeta^i(\widetilde{\mathbf{x}}_k^i)$ summarizes all information flowing from the object detection variable \(D_k^i\) and the consistency factors \(\Phi_k^{i,j}\) back to the Bernoulli state node $\widetilde{\mathbf{x}}_k^i$.

To compute the messages from the detection subgraph to \(\alpha_k^{i,j}\), we first form the extrinsic messages from \(D_k^i\) to \(\Phi_k^{i,j}\),
\begin{equation}
\label{eq_D_to_phi}
\nu^{i,j}\left(D_k^i\right)
=
\prod_{\substack{l=1\\l\neq j}}^{m_k}{\mu}^{i,l}\left(D_k^i\right) \int
f_k^{D,i} \left(D_k^i | \widetilde{\mathbf{x}}_k^i\right)
\chi^{i} \left(\widetilde{\mathbf{x}}_k^i\right)
 \delta \widetilde{\mathbf{x}}_k^i,
\end{equation}
where $\chi^{i}$ is the message from $\widetilde{\mathbf{x}}_k^i$ to $f_k^{D,i}$, given by
\begin{equation}
  \label{eq_chi}
  \chi^{i}\left(\widetilde{\mathbf{x}}_k^i\right)
=
\underline{f}_{k|k-1}^{i}\left(\widetilde{\mathbf{x}}_k^i\right)
\prod_{j=1}^{m_k}
\varrho^{i,j}\left(\widetilde{\mathbf{x}}_k^i\right).
\end{equation}
Using the definition of $f_k^{D,i}$ \eqref{eq_detection_def}, we obtain
\begin{align}
\label{eq_nu_closed}
\nu^{i,j}(1)
&= 
\prod_{\substack{l=1\\l\neq j}}^{m_k}{\mu}^{i,l}(1)
\int
p^D_k(x)e^{-\gamma_k(x)}
\chi^{i}(\{(i,x)\})dx,
\\
\nu^{i,j}(0) &=  \prod_{\substack{l=1\\l\neq j}}^{m_k}{\mu}^{i,l}(0)
\left[\int
\left(1-p^D_k(x)\right)
\chi^{i}\big(\{(i,x)\}\big)dx
+
\chi^{i}(\emptyset)\right].
\end{align}
Finally, the message from \(\Phi_k^{i,j}\) to \(\alpha_k^{i,j}\) is
\begin{align}
\label{eq_lambda_def}
\lambda^{i,j}\left(\alpha_k^{i,j}\right)
&= 
\sum_{D_k^i\in\{0,1\}}
\Phi_k^{i,j}\left(D_k^i,\alpha_k^{i,j}\right)\nu^{i,j}\left(D_k^i\right).
\end{align}
Using \eqref{binary_factor_detection} and normalizing $\lambda^{i,j}(\alpha_k^{i,j})$ by $\lambda^{i,j}(0)$, we obtain $\overline{\lambda}^{i,j}(0) = 1$ and 
\begin{equation}
\label{eq_lambda_closed}
\overline{\lambda}^{i,j}(1)
=
\frac{\nu^{i,j}(1)}{\nu^{i,j}(0)+\nu^{i,j}(1)}.
\end{equation}
Here, the normalized message \(\overline{\lambda}^{i,j}(\alpha_k^{i,j})\) quantifies, for the \(i\)-th existing Bernoulli component, the compatibility between the detection event \(D_k^i\) and the assignment of measurement \(z_k^j\) to that component. 

\subsection{Messages through the association consistency factor}

For each measurement $z_k^j$, \(j\in\{1,\dots,m_k\}\), we denote the admissible support of \(\beta_k^j\) as $\mathcal I_j = \{1,\dots,n_{k|k-1},n_{k|k-1}+j,\dots,n_{k|k}\}$. The incoming extrinsic message from $\alpha_k^{i,j}$ to \(\Psi_k^{i,j}\) is obtained by combining the messages passed from the local likelihood factor and the object detection consistency factor. Specifically, for an existing Bernoulli component $i\in\{1,\dots,n_{k|k-1}\}$, with $j\in\{1,\dots,m_k\}$,
\begin{equation}
\label{eq_eta_alpha}
\eta_{\Psi}^{i,j}\left(\alpha_k^{i,j}\right)
=
\overline{\vartheta}^{i,j}\left(\alpha_k^{i,j}\right)
\overline{\lambda}^{i,j}\left(\alpha_k^{i,j}\right).
\end{equation}
For a new Bernoulli component \(i=n_{k|k-1}+j'\), with \(j'\in\{1,\dots,m_k\}\), $j\in\{1,\dots,j^\prime\}$,
\begin{equation}
\label{eq_eta_alpha_new}
\eta_{\Psi}^{i,j}\left(\alpha_k^{i,j}\right)
=
\overline{\vartheta}^{i,j}\left(\alpha_k^{i,j}\right).
\end{equation}
The message from \(\Psi_k^{i,j}\) to \(\beta_k^j\) is
\begin{equation}
\label{eq_kappa_def}
\kappa^{i,j}\left(\beta_k^j\right)
= 
\sum_{\alpha_k^{i,j}\in\{0,1\}}
\Psi_k^{i,j}\left(\alpha_k^{i,j},\beta_k^j\right)
\eta_{\Psi}^{i,j}\left(\alpha_k^{i,j}\right) =
\begin{cases}
\eta_{\Psi}^{i,j}(1) & \beta_k^j=i\\
1 & \beta_k^j\neq i.
\end{cases}
\end{equation}
The message from \(\beta_k^j\) back to \(\Psi_k^{i,j}\) is given by the product of all incoming messages to \(\beta_k^j\), excluding the one from \(\Psi_k^{i,j}\),
\begin{equation}
\label{eq_beta_to_psi}
\rho^{i,j}\left(\beta_k^j\right)
= 
\prod_{l\in\mathcal I_j \setminus \{i\}}
{\kappa}^{l,j}\left(\beta_k^j\right) = \begin{cases}
  1 & \beta_k^j = i \\
 \eta_{\Psi}^{\beta_k^j,j}(1) & \beta_k^j \in \mathcal{I}_j \setminus \{i\}.
\end{cases}
\end{equation}
The message from \(\Psi_k^{i,j}\) to \(\alpha_k^{i,j}\) is then
\begin{equation}
\label{eq_xi_def}
\xi^{i,j}\left(\alpha_k^{i,j}\right)
= 
\sum_{\beta_k^j\in\mathcal I_j}
\Psi_k^{i,j}\left(\alpha_k^{i,j},\beta_k^j\right)
\rho^{i,j}\left(\beta_k^j\right) = \begin{cases}
  1 & \alpha_k^{i,j} = 1 \\
  \sum_{l \in \mathcal{I}_j \setminus \{i\}} \eta_{\Psi}^{l,j}(1) & \alpha_k^{i,j} = 0.
\end{cases}
\end{equation}
Here, the messages \({\kappa}^{i,j}(\beta_k^j)\) and \({\xi}^{i,j}(\alpha_k^{i,j})\) quantify the local agreement enforced by the consistency factor \(\Psi_k^{i,j}\) between the dual data association representations.

\subsection{Messages from local likelihood factor node to Bernoulli state node}

To compute the message \(\varrho^{i,j}(\widetilde{\mathbf{x}}_k^i)\) from the \(j\)-th local likelihood branch back to \(\widetilde{\mathbf{x}}_k^i\), we first combine the incoming messages at $\alpha_k^{i,j}$. Specifically, for existing Bernoulli component $i\in\{1,\dots,n_{k|k-1}\}$, with $j\in\{1,\dots,m_k\}$,
\begin{equation}
\label{eq_tau_def}
\tau^{i,j}\left(\alpha_k^{i,j}\right)
=
\overline{\lambda}^{i,j}\left(\alpha_k^{i,j}\right)
{\xi}^{i,j}\left(\alpha_k^{i,j}\right).
\end{equation}
For a new Bernoulli component \(i=n_{k|k-1}+j'\), with \(j'\in\{1,\dots,m_k\}\), $j\in\{1,\dots,j^\prime\}$,
\begin{equation}
\label{eq_tau_new}
\tau^{i,j}\left(\alpha_k^{i,j}\right)
=
{\xi}^{i,j}\left(\alpha_k^{i,j}\right).
\end{equation}
Then, for factors of type \(\underline{s}_k^i(\widetilde{\mathbf{x}}_k^i,\alpha_k^{i,j};z_k^j)\), we have
\begin{equation}
\label{eq_rho_exist}
\varrho^{i,j}\left(\widetilde{\mathbf{x}}_k^i\right)
= 
\sum_{\alpha_k^{i,j}\in\{0,1\}}
\underline{s}_k^i\left(\widetilde{\mathbf{x}}_k^i,\alpha_k^{i,j};z_k^j\right)
\tau^{i,j}\left(\alpha_k^{i,j}\right) =
\begin{cases}
\delta_i[u]\gamma_k(x)\ell_k\left(z_k^j | x\right)\tau^{i,j}(1)+\tau^{i,j}(0),
& \widetilde{\mathbf{x}}_k^i=\{(u,x)\}\\
\tau^{i,j}(0)
& \widetilde{\mathbf{x}}_k^i=\emptyset\\
0 & \text{otherwise}.
\end{cases}
\end{equation}
For factor \(\overline{s}_k^i(\widetilde{\mathbf{x}}_k^i,\alpha_k^{i,j};z_k^j)\) of a new Bernoulli component \(i=n_{k|k-1}+j\), we similarly obtain
\begin{equation}
\label{eq_rho_new}
\varrho^{i,j}\left(\widetilde{\mathbf{x}}_k^i\right)
= 
\sum_{\alpha_k^{i,j}\in\{0,1\}}
\overline{s}_k^i\left(\widetilde{\mathbf{x}}_k^i,\alpha_k^{i,j};z_k^j\right)
\tau^{i,j}\left(\alpha_k^{i,j}\right) =
\begin{cases}
\delta_i[u]\gamma_k(x)\ell_k\left(z_k^j | x\right)\tau^{i,j}(1)
& \widetilde{\mathbf{x}}_k^i=\{(u,x)\}\\
\lambda_k^C\left(z_k^j\right)\tau^{i,j}(1)+\tau^{i,j}(0)
& \widetilde{\mathbf{x}}_k^i=\emptyset\\
0, & \text{otherwise}.
\end{cases}
\end{equation}
Therefore, the message $\varrho^{i,j}(\widetilde{\mathbf{x}}_k^i)$ summarizes the contribution of the $j$-th local likelihood branch to the Bernoulli state variable $\widetilde{\mathbf{x}}_k^i$, after combining the local measurement compatibility with the association-side consistency information.

\subsection{Belief calculations}

The belief of each Bernoulli state variable \(\widetilde{\mathbf{x}}_k^i\) is obtained by multiplying all incoming messages, together with the corresponding prior factor. For an existing Bernoulli component \(i\in\{1,\dots,n_{k|k-1}\}\), the belief is
\begin{align}
\label{eq_belief_exist}
b^i\left(\widetilde{\mathbf{x}}_k^i\right)
&\propto 
\underline{f}_{k|k-1}^{i}\left(\widetilde{\mathbf{x}}_k^i\right)
\zeta^i\left(\widetilde{\mathbf{x}}_k^i\right)
\prod_{j=1}^{m_k}
\varrho^{i,j}\left(\widetilde{\mathbf{x}}_k^i\right),
\end{align}
where \(\zeta^i(\widetilde{\mathbf{x}}_k^i)\) \eqref{eq_zeta_closed} is the collapsed message from the detection subgraph.
For a new Bernoulli component \(i=n_{k|k-1}+j'\), with \(j'\in\{1,\dots,m_k\}\), the belief is
\begin{align}
\label{eq_belief_new}
b^i\left(\widetilde{\mathbf{x}}_k^i\right)
&\propto 
\overline{f}_{k|k-1}^{i}\left(\widetilde{\mathbf{x}}_k^i\right)
\prod_{j=1}^{j'}
\varrho^{i,j}\left(\widetilde{\mathbf{x}}_k^i\right).
\end{align}

After normalization, the beliefs in \eqref{eq_belief_exist} and \eqref{eq_belief_new} provide the loopy BP approximation of the marginal Bernoulli densities \(f_{k|k}^i(\widetilde{\mathbf{x}}_k^i)\), \(i\in\{1,\dots,n_{k|k}\}\). In particular, the approximate existence probability and conditional single object density of the \(i\)-th Bernoulli component can be recovered directly from the corresponding belief \(b^i(\widetilde{\mathbf{x}}_k^i)\).

\subsection{Initialization and pseudocode}
To initialize loopy BP, we can first compute the extrinsic messages \eqref{eq_epsilon_exist} and \eqref{eq_epsilon_new} from Bernoulli state variables $\widetilde{\mathbf{x}}_k^i$ to local likelihood factors, $\underline{s}_k^i$ or $\overline{s}_k^i$, according to the case of misdetection. Specifically,
for an existing Bernoulli component $i\in\{1,\dots,n_{k|k-1}\}$, such initialization is induced by the prior factor \eqref{eq_prior_exist}, together with the no-association branch in the ZIP model  \eqref{eq_detection_def}, for all $j\in\{1,\dots,m_k\}$, which is given by
\begin{equation}
  \label{eq_extrinsic_exist_ini}
  \epsilon^{i,j}\left(\widetilde{\mathbf{x}}_k^i\right)
=
\underline{f}_{k|k-1}^{i}\left(\widetilde{\mathbf{x}}_k^i\right) \sum_{D_k^i\in\{0,1\}}f_k^{D,i}\left(D_k^i | \widetilde{\mathbf{x}}_k^i\right).
\end{equation} 
As for new Bernoulli components, the initialization is induced by the prior factor \eqref{eq_prior_new}. 
In addition, the message $\chi^i$ from the Bernoulli state variable $\widetilde{\mathbf{x}}_k^i$ to the factor $f_k^{D,i}$ is initialized as $\chi^i(\widetilde{\mathbf{x}}_k^i) = \underline{f}^i_{k|k-1}(\widetilde{\mathbf{x}}_k^i)$, cf. \eqref{eq_chi}. 

With these initialized messages in place, the remaining messages are computed from the corresponding recursions, and all the messages are then updated in parallel. The iterations are typically repeated for a fixed number of iterations \cite{florian2021scalable,xia2023trajectory}. After termination, the beliefs $b^i(\widetilde{\mathbf{x}}_k^i)$, $i\in\{1,\dots,n_{k|k}\}$, provide the approximate marginal Bernoulli densities.

The complete loopy BP update derived above is summarized in Algorithm \ref{alg:lbp_zip}. The operations within each message update stage can be evaluated in parallel over the corresponding Bernoulli component and measurement indices. In the particle implementation presented in the next section, the state-dependent messages in Algorithm~\ref{alg:lbp_zip} are represented by an existence probability and weighted particles, whereas the association and detection messages are scalar.

\begin{algorithm}[h!]
\caption{Loopy BP update for the extended object PMB filter under the ZIP measurement model}
\label{alg:lbp_zip}

\KwIn{Predicted Bernoulli factors
\(\{\underline{f}_{k|k-1}^{i}\}_{i=1}^{n_{k|k-1}}\);
new Bernoulli factors
\(\{\overline{f}_{k|k-1}^{i}\}_{i=n_{k|k-1}+1}^{n_{k|k}}\);
measurement set \(\mathbf z_k\);
maximum number of iterations \(P_{\max}\).}

\KwOut{Approximate marginal Bernoulli densities
\(\{f_{k|k}^{i}\}_{i=1}^{n_{k|k}}\).}

Set \(n_{k|k}=n_{k|k-1}+m_k\)\;

Initialize the extrinsic state messages
\(\epsilon^{i,j,(1)}\) for the existing Bernoulli components using
\eqref{eq_extrinsic_exist_ini}\;

Initialize the extrinsic state messages
\(\epsilon^{i,j,(1)}\) for the new Bernoulli components using
their prior factors \(\overline{f}_{k|k-1}^{i}\)\;

Initialize
\(\chi^{i,(1)}=\underline{f}_{k|k-1}^{i}\),
\(i=1,\ldots,n_{k|k-1}\)\;

Initialize all remaining scalar messages to one\;

Set \(p\leftarrow 0\);

\Repeat{
\(p=P_{\max}\)
}{
    Set \(p\leftarrow p+1\)\;

    Compute the local likelihood-to-association messages
    \(\overline{\vartheta}^{i,j,(p)}\) using
    \eqref{eq_theta_exist} and \eqref{eq_theta_new}\;

    Update the detection subgraph messages
    \(\mu^{i,j,(p)}\), \(\zeta^{i,(p)}\),
    \(\nu^{i,j,(p)}\), and
    \(\overline{\lambda}^{i,j,(p)}\) using
    \eqref{eq_phi_to_D}, \eqref{eq_zeta_closed},
    \eqref{eq_nu_closed}, and \eqref{eq_lambda_closed}\;

    Update the association consistency messages
    \(\eta_{\Psi}^{i,j,(p)}\),
    \(\kappa^{i,j,(p)}\),
    \(\rho^{i,j,(p)}\), and
    \(\xi^{i,j,(p)}\) using
    \eqref{eq_eta_alpha}--\eqref{eq_xi_def}\;

    Compute the messages
    \(\tau^{i,j,(p)}\) and
    \(\varrho^{i,j,(p)}\) from the local likelihood branches using
    \eqref{eq_tau_def}--\eqref{eq_rho_new}\;

    Update and normalize
    \(\epsilon^{i,j,(p+1)}\) and
    \(\chi^{i,(p+1)}\) using
    \eqref{eq_epsilon_exist}, \eqref{eq_epsilon_new}, and
    \eqref{eq_chi}\;
}

Compute and normalize the final beliefs
\(b^i(\widetilde{\mathbf{x}}_k^i)\) using
\eqref{eq_belief_exist} and \eqref{eq_belief_new}\;

Set
\(f_{k|k}^{i}(\widetilde{\mathbf{x}}_k^i)
=b^i(\widetilde{\mathbf{x}}_k^i)\),
\(i=1,\ldots,n_{k|k}\)\;

\Return{\(\{f_{k|k}^{i}\}_{i=1}^{n_{k|k}}\)}\;
\end{algorithm}

\section{Particle Implementation}
\label{sec_particle}
In this section, we present a particle-based loopy BP implementation of the extended object PMB filter under the ZIP measurement model, where each Bernoulli component is represented by an existence probability and a particle representation of its single object density. We first describe the prediction step, and then present the implementation of the loopy BP update derived in Section \ref{sec_lbp}. We also discuss several practical aspects of the implementation.

We represent the PMB density at time step \(k^\prime\in\{k-1,k\}\) using a hybrid form. The PPP intensity for undetected objects is kept in analytic form as \(\lambda_{k|k^\prime}(x)\), whereas each Bernoulli component is represented by its existence probability \(r_{k|k^\prime}^i\) and a particle approximation of its single object density,
\begin{equation}
\label{eq_particle_ber}
f_{k|k^\prime}^i(x)
=
\sum_{l=1}^{L}
w_{k|k^\prime}^{i,(l)}
\delta_{x_{k|k^\prime}^{i,(l)}}(x),
\end{equation}
where $L$ is the number of particles, \(x_{k|k^\prime}^{i,(l)}\) is the $l$-th particle, \(w_{k|k^\prime}^{i,(l)}\) is its normalized weight such that \(\sum_{l=1}^{L} w_{k|k^\prime}^{i,(l)} = 1\). In what follows, we omit the auxiliary variable $u$ in the object state for simplicity.

\subsection{Prediction Step}
\label{sec_particle_predict}
The Poisson intensity for undetected objects is predicted analytically according to \eqref{eq_ppp_predict} in Lemma \ref{pmbm_prediction}. In contrast, the Bernoulli components are propagated over time by particles. Specifically, let the \(i\)-th Bernoulli component, with $i\in \{1,\dots,n_{k-1|k-1}\}$, at time step \(k-1\) be represented by
\begin{equation}
\left\{r_{k-1|k-1}^i,
\left(w_{k-1|k-1}^{i,(l)},x_{k-1|k-1}^{i,(l)}\right)_{l=1}^{L}\right\}.
\end{equation}
For each particle \(x_{k-1|k-1}^{i,(l)}\), to compute its predicted state at time step \(k\), we draw a predicted particle \(x_{k|k-1}^{i,(l)}\) from a proposal density $q_{k|k-1}^i(\cdot| x_{k-1|k-1}^{i,(l)})$, and its corresponding unnormalized importance weight is given by
\begin{equation}
\label{eq_importance_weight_predict}
\widetilde{w}_{k|k-1}^{i,(l)}
=
w_{k-1|k-1}^{i,(l)}
\frac{
p_k^S\left(x_{k-1|k-1}^{i,(l)}\right)
g_k\left(x_{k|k-1}^{i,(l)} | x_{k-1|k-1}^{i,(l)}\right)
}{
q_{k|k-1}^i\left(x_{k|k-1}^{i,(l)} | x_{k-1|k-1}^{i,(l)}\right)
},
\end{equation}
and is normalized as
\begin{equation}
\label{eq_predict_weight_norm}
w_{k|k-1}^{i,(l)}
=
\frac{\widetilde{w}_{k|k-1}^{i,(l)}}
{\sum_{l^\prime=1}^{L}\widetilde{w}_{k|k-1}^{i,(l^\prime)}}.
\end{equation}
The predicted existence probability is given by
\begin{equation}
\label{eq_predict_exist_particle}
r_{k|k-1}^i
=
r_{k-1|k-1}^i
\sum_{l=1}^{L}
w_{k-1|k-1}^{i,(l)}
p_k^S\left(x_{k-1|k-1}^{i,(l)}\right).
\end{equation}
The $i$-th predicted Bernoulli density, with $i\in \{1,\dots,n_{k|k-1}\}$, is then represented by 
\begin{equation} \label{eq_initial_extrinsic_exist}
  \left\{r_{k|k-1}^i,
\left(w_{k|k-1}^{i,(l)},x_{k|k-1}^{i,(l)}\right)_{l=1}^{L}\right\},
\end{equation}
where $n_{k|k-1} = n_{k-1|k-1}$.

\subsection{Update Step}

We now describe the particle implementation of the update step that includes initialization, sum-product message passing, and belief calculation. We note that in each iteration of loopy BP, the particle states remain unchanged, but their weights are refined. For the (normalized) extrinsic message $\epsilon^{i,j,(p)}(\widetilde{\mathbf{x}}_k^i)$ at the $p$-th iteration, it is parameterized by 
\begin{equation} 
  \left\{r_{k,\epsilon}^{i,j,(p)},
\left(w_{k,\epsilon}^{i,j,(l),(p)},x_{k|k}^{i,(l)}\right)_{l=1}^{L}\right\},
\end{equation}
and the (normalized) message $\chi^i(\widetilde{\mathbf{x}}_k^i)$ at the $p$-th iteration is parameterized by 
\begin{equation} 
  \label{eq_initial_particle_chi}
  \left\{r_{k,\chi}^{i,(p)},
\left(w_{k,\chi}^{i,(l),(p)},x_{k|k}^{i,(l)}\right)_{l=1}^{L}\right\}.
\end{equation}

\subsubsection{Initialization}

For the $i$-th existing Bernoulli component, its extrinsic messages $\epsilon^{i,j,(1)}(\widetilde{\mathbf{x}}_k^i)$, $j\in\{1,\dots,m_k\}$ at the first iteration are initialized according to \eqref{eq_extrinsic_exist_ini}. Specifically, we have $x_{k|k}^{i,(l)} = x_{k|k-1}^{i,(l)}$ for $l\in\{1,\dots,L\}$, and for each particle $x_{k|k}^{i,(l)}$, its unnormalized weight is 
\begin{equation}
\widetilde{w}_{k,\epsilon}^{i,j,(l),(1)} =
w_{k|k-1}^{i,(l)}
  \left[1 - p_k^D\left(x_{k|k}^{i,(l)}\right) + p_k^D\left(x_{k|k}^{i,(l)}\right)
  e^{-\gamma_k\left(x_{k|k}^{i,(l)}\right)}\right],
\end{equation}
which are identical for all $j\in\{1,\dots,m_k\}$ and normalized as
\begin{equation}
\label{eq_weight_new_norm}
w_{k,\epsilon}^{i,j,(l),(1)}
=
\frac{\widetilde{w}_{k,\epsilon}^{i,j,(l),(1)}}
     {\sum_{l^\prime=1}^{L}\widetilde{w}_{k,\epsilon}^{i,j,(l^\prime),(1)}}.
\end{equation}
The initial existence probability of the \(i\)-th existing Bernoulli component is given by
\begin{equation}
\label{eq_exist_exist_init}
r_{k,\epsilon}^{i,j,(1)}
=
\frac{
  r_{k|k-1}^i\sum_{l=1}^{L}\widetilde{w}_{k,\epsilon}^{i,j,(l),(1)}
}{
  1-r_{k|k-1}^i+r_{k|k-1}^i\sum_{l=1}^{L}\widetilde{w}_{k,\epsilon}^{i,j,(l),(1)}
}.
\end{equation}
In addition, the message $\chi^i(\widetilde{\mathbf{x}}_k^i)$ \eqref{eq_initial_particle_chi} is initialized as \eqref{eq_initial_extrinsic_exist}.

As for a new Bernoulli component \(i=n_{k|k-1}+j^\prime\), where \(j^\prime\in\{1,\dots,m_k\}\), to initialize its extrinsic messages $\epsilon^{i,j,(1)}(\widetilde{\mathbf{x}}_k^i)$ for \(j\in\{1,\dots,j'\}\), we first draw the particles
$
\{x_{k|k}^{i,(l)}\}_{l=1}^{L}
$
from a proposal density $\overline q_k^i(x)$. According to \eqref{eq_prior_new}, the corresponding unnormalized importance weights are
\begin{equation}
\label{eq_importance_weight_new}
\widetilde{w}_{k,\epsilon}^{i,j,(l),(1)}
=
\frac{
  p_k^D\left(x_{k|k}^{i,(l)}\right)
  e^{-\gamma_k\left(x_{k|k}^{i,(l)}\right)}
  \lambda_{k|k-1}\left(x_{k|k}^{i,(l)}\right)
}{
  \overline q_k^i\left(x_{k|k}^{i,(l)}\right)
},
\end{equation}
which are identical for all valid $j$ and normalized as in \eqref{eq_weight_new_norm}, and the initial existence probability of the \(i\)-th new Bernoulli component is given by
\begin{equation}
\label{eq_exist_new_init}
r_{k,\epsilon}^{i,j,(1)}
=
\frac{
  \sum_{l=1}^{L}\widetilde{w}_{k,\epsilon}^{i,j,(l),(1)}
}{
  1+\sum_{l=1}^{L}\widetilde{w}_{k,\epsilon}^{i,j,(l),(1)}
}.
\end{equation}

\subsubsection{Sum-product message passing}

For an existing Bernoulli component \(i\in\{1,\dots,n_{k|k-1}\}\), the message from $\underline{s}_k^i$ to $\alpha_k^{i,j}$ \eqref{eq_theta_exist} for $j\in\{1,\dots,m_k\}$ is given by
\begin{equation}
\label{eq_theta_exist_particle}
\overline{\vartheta}^{i,j,(p)}(1)
\approx r_{k,\epsilon}^{i,j,(p)}
\sum_{l=1}^{L}
w_{k,\epsilon}^{i,j,(l),(p)}
\gamma_k\left(x_{k|k}^{i,(l)}\right)
\ell_k\left(z_k^j | x_{k|k}^{i,(l)}\right).
\end{equation}

For a new Bernoulli component \(i=n_{k|k-1}+j^\prime\), with \(j^\prime\in\{1,\dots,m_k\}\) and $j < j^\prime$, the message from $\underline{s}_k^i$ to $\alpha_k^{i,j}$ \eqref{eq_theta_exist} is also given by \eqref{eq_theta_exist_particle}. The message from $\overline{s}_k^i$ to $\alpha_k^{i,j}$ \eqref{eq_theta_new} when $j = j^\prime$ is given by
\begin{equation}
\label{eq_theta_new_particle}
\overline{\vartheta}^{i,j,(p)}(1)
\approx
\frac{r_{k,\epsilon}^{i,j,(p)}
\sum_{l=1}^{L}
w_{k,\epsilon}^{i,j,(l),(p)}
\gamma_k\left(x_{k|k}^{i,(l)}\right)
\ell_k\left(z_k^j | x_{k|k}^{i,(l)}\right)
}{
1 - r_{k,\epsilon}^{i,j,(p)}
}
+
\lambda_k^C(z_k^j).
\end{equation}

For an existing Bernoulli component, the message from \(D_k^i\) to \(\Phi_k^{i,j}\) \eqref{eq_nu_closed} is given by
\begin{align}
\label{eq_nu_particle}
\nu^{i,j,(p)}(1) &= 
\prod_{\substack{j^\prime=1\\j^\prime\neq j}}^{m_k}\mu^{i,j^\prime,(p)}(1) \left[ r_{k,\chi}^{i,(p)}
\sum_{l=1}^{L}
w_{k,\chi}^{i,(l),(p)}
p_k^D\left(x_{k|k}^{i,(l)}\right)
e^{-\gamma_k\left(x_{k|k}^{i,(l)}\right)} \right],
\\
\nu^{i,j,(p)}(0) &= 
\prod_{\substack{j^\prime=1\\j^\prime\neq j}}^{m_k}\mu^{i,j^\prime,(p)}(0) 
\left[ 1 - r_{k,\chi}^{i,(p)}
\sum_{l=1}^{L}
w_{k,\chi}^{i,(l),(p)}
p_k^D\left(x_{k|k}^{i,(l)}\right)
\right].
\end{align}

The remaining messages \(\mu^{i,j}\), \(\lambda^{i,j}\), \(\kappa^{i,j}\), \(\rho^{i,j}\), \(\xi^{i,j}\), and \(\tau^{i,j}\) are all scalar and are updated exactly according to the recursions in Section~\ref{sec_lbp}. Also, according to \eqref{eq_zeta_closed}, the message $\zeta^i (\widetilde{\mathbf{x}}_k^i)$ can be evaluated on each particle \(x_{k|k}^{i,(l)}\) using the scalar message $\mu^{i,j}$. Similarly, using \eqref{eq_rho_exist} or \eqref{eq_rho_new}, the message \(\varrho^{i,j}(\widetilde{\mathbf{x}}_k^i)\) can be evaluated on each particle \(x_{k|k}^{i,(l)}\) using the scalar message $\tau^{i,j}$. Consequently, 
the updated extrinsic messages $\epsilon^{i,j,(p+1)}(\widetilde{\mathbf{x}}_k^i)$ at the next iteration are represented on the same particle support by updated particle weights for $i\in\{1,\dots,n_{k|k}\}$ using \eqref{eq_epsilon_exist} or \eqref{eq_epsilon_new}. Specifically, for the case of \eqref{eq_epsilon_exist} with index pair $(i,j)$, with $i\in\{1,\dots,n_{k|k-1}\}$, $j\in\{1,\dots,m_k\}$, the unnormalized weight of particle $x^{i,(l)}$ is 
\begin{equation}
\label{eq_extrinsic_weight_exist}
\widetilde{w}_{k,\epsilon}^{i,j,(l),(p+1)} = 
w_{k|k-1}^{i,(l)}
\zeta^{i,(p)}\left(\{(i,x_{k|k}^{i,(l)})\}\right)
\prod_{\substack{l^\prime=1 \\ l^\prime\neq j}}^{m_k}
\varrho^{i,l^\prime,(p)}\left(\{(i,x_{k|k}^{i,(l)})\}\right),
\end{equation}
which is normalized as
\begin{equation}
  \label{eq_normalize_weight_extrinsic}
  w_{k,\epsilon}^{i,j,(l),(p+1)} = \frac{\widetilde{w}_{k,\epsilon}^{i,j,(l),(p+1)}}{\sum_{l^\prime=1}^L \widetilde{w}_{k,\epsilon}^{i,j,(l^\prime),(p+1)}}.
\end{equation}
When $\epsilon^{i,j,(p+1)}(\widetilde{\mathbf{x}}_k^i)$ is evaluated at an empty set, we have 
\begin{equation}
  \label{eq_extrinsic_weight_exist_empty}
  \widetilde{w}^{i,j,(p+1)}_\emptyset = \left(1-r^i_{k|k-1}\right)\zeta^{i,(p)}(\emptyset)
\prod_{\substack{l^\prime=1 \\ l^\prime\neq j}}^{m_k}
\varrho^{i,l^\prime,(p)}(\emptyset).
\end{equation}
After normalization, the existence probability of the extrinsic message $\epsilon^{i,j,(p+1)}(\widetilde{\mathbf{x}}_k^i)$ \eqref{eq_epsilon_exist} is given by 
\begin{equation}
  \label{eq_normalize_exist_extrinsic}
  r_{k,\epsilon}^{i,j,(p+1)} = \frac{r^i_{k|k-1}\sum_{l=1}^L \widetilde{w}_{k,\epsilon}^{i,j,(l),(p+1)}}{r^i_{k|k-1}\sum_{l=1}^L \widetilde{w}_{k,\epsilon}^{i,j,(l),(p+1)} + \widetilde{w}^{i,j,(p+1)}_\emptyset}.
\end{equation}

For the case of \eqref{eq_epsilon_new} with index pair $(i,j)$, $i = n_{k|k-1}+j^\prime$, $j^\prime\in\{1,\dots,m_k\}$, and $j\in\{1,\dots,j^\prime\}$, the unnormalized weight of particle $x_{k|k}^{i,(l)}$ is 
\begin{equation}
\label{eq_extrinsic_weight_new}
\widetilde{w}_{k,\epsilon}^{i,j,(l),(p+1)}
=
w_{k,\epsilon}^{i,j,(l),(1)}
\prod_{\substack{l^\prime=1 \\ l^\prime\neq j}}^{j^\prime}
\varrho^{i,l^\prime,(p)}\left(\{(i,x_{k|k}^{i,(l)})\}\right),
\end{equation}
with $w_{k,\epsilon}^{i,j,(l),(1)}$ given in \eqref{eq_weight_new_norm}, and its normalized weight is also computed as in \eqref{eq_normalize_weight_extrinsic}. Similarly, when evaluated at an empty set, we have 
\begin{equation}
  \label{eq_extrinsic_weight_new_empty}
  \widetilde{w}^{i,j,(p+1)}_\emptyset = \left(1-r_{k,\epsilon}^{i,j,(1)}\right)
\prod_{\substack{l^\prime=1 \\ l^\prime\neq j}}^{j^\prime}
\varrho^{i,l^\prime,(p)}(\emptyset),
\end{equation}
with $r_{k,\epsilon}^{i,j,(1)}$ specified in \eqref{eq_exist_new_init}, and the existence probability of the extrinsic message $\epsilon^{i,j,(p+1)}(\widetilde{\mathbf{x}}_k^i)$ \eqref{eq_epsilon_new} after normalization is also computed as in \eqref{eq_normalize_exist_extrinsic}, but with $r^i_{k|k-1}$ replaced by $r_{k,\epsilon}^{i,j,(1)}$.

In addition, for the $i$-th existing Bernoulli component with $i\in\{1,\dots,n_{k|k-1}\}$, the message $\chi^i(\widetilde{\mathbf{x}}_k^i)$ in the detection subgraph from $\widetilde{\mathbf{x}}_k^i$ to $f_k^{D,i}$ at the next iteration is also represented using the same particle support by updated particle weights. According to \eqref{eq_chi}, the unnormalized weight of particle $x^{i,(l)}$ is 
\begin{equation}
\widetilde{w}_{k,\chi}^{i,(l),(p+1)} =
w_{k|k-1}^{i,(l)}
\prod_{j=1}^{m_k}
\varrho^{i,j,(p)}\left(\{(i,x_{k|k}^{i,(l)})\}\right),
\end{equation}
which is normalized as
\begin{equation}
  w_{k,\chi}^{i,(l),(p+1)} = \frac{\widetilde{w}_{k,\chi}^{i,(l),(p+1)}}{\sum_{l^\prime=1}^L \widetilde{w}_{k,\chi}^{i,(l^\prime),(p+1)}}.
\end{equation}
When $\chi^{i,(p+1)}(\widetilde{\mathbf{x}}_k^i)$ is evaluated at an empty set, we have 
\begin{equation}
  \widetilde{w}^{i,(p+1)}_\emptyset = \left(1-r^i_{k|k-1}\right)
\prod_{\substack{j=1}}^{m_k}
\varrho^{i,j,(p)}(\emptyset).
\end{equation}
After normalization, the existence probability of the message $\chi^{i,(p+1)}(\widetilde{\mathbf{x}}_k^i)$ \eqref{eq_chi} is given by 
\begin{equation}
  r_{k,\chi}^{i,(p+1)} = \frac{r^i_{k|k-1}\sum_{l=1}^L \widetilde{w}_{k,\chi}^{i,(l),(p+1)}}{r^i_{k|k-1}\sum_{l=1}^L \widetilde{w}_{k,\chi}^{i,(l),(p+1)} + \widetilde{w}^{i,(p+1)}_\emptyset}.
\end{equation}

\subsubsection{Belief calculation}

At the last iteration, we compute the beliefs $b^i(\widetilde{\mathbf{x}}_k^i)$ for $i\in\{1,\dots,n_{k|k}\}$. Each belief is a Bernoulli set density, which can be calculated similarly to the extrinsic messages
$\epsilon^{i,j,(p)}(\widetilde{\mathbf{x}}_k^i)$ with the difference that the products in \eqref{eq_extrinsic_weight_exist}, \eqref{eq_extrinsic_weight_exist_empty}, \eqref{eq_extrinsic_weight_new} and \eqref{eq_extrinsic_weight_new_empty}
now need to enumerate every element. 

\subsection{Practical Considerations}

First, we comment on the convergence of loopy BP. Since the factor graph in Fig.~\ref{fig_factor_graph} contains cycles, there is in general no theoretical guarantee that the parallel message passing recursions converge. Nevertheless, many empirical results have shown that loopy BP for multiple EOT converge within few iterations \cite{florian2021scalable,xia2023trajectory}. In particular, the proposed loopy BP update achieves reasonable performance with only two iterations, and further improvements largely saturate by five iterations; see Table~\ref{tab:pmb_bp_zip_ablation} in Section~\ref{sec_simulation}.

Next, we analyze the computational complexity of the particle-BP update. Let \(L\) and \(P_{\max}\) denote the numbers of particles per Bernoulli component and BP iterations, respectively. Existing Bernoulli components contribute \(n_{k|k-1}m_k\) component-measurement pairs, whereas the measurement-initiated components contribute \(\sum_{j=1}^{m_k}j=m_k(m_k+1)/2\). Thus, the total number of association edges is
\begin{equation}
  E_k=n_{k|k-1}m_k+\frac{m_k(m_k+1)}{2},
\end{equation}
and for each edge, the particle sums and particle-wise message updates require \(\mathcal{O}(L)\) operations, while the remaining association messages are scalar. By reusing the common sums and products in the leave-one-out message updates, all messages in one BP iteration can therefore be computed in \(\mathcal{O}(LE_k)\) time. Initialization and belief calculation have the same order and do not change the dominant scaling, yielding
\begin{equation}
\mathcal{O}\!\left(P_{\max}L
\left(n_{k|k-1}m_k+m_k^2\right)\right).
\end{equation}
For fixed \(L\) and \(P_{\max}\), this reduces to \(\mathcal{O}(n_{k|k-1}m_k+m_k^2)=\mathcal{O}(n_{k|k}m_k)\). Bernoulli prediction and new-particle generation add \(\mathcal{O}(L(n_{k-1|k-1}+m_k))\). since each measurement initiates a new Bernoulli component with $n_{k|k} = n_{k|k-1}+m_k$, and an extended object typically generates multiple measurements, the overall computational complexity can conservatively be regarded as quadratic in the number of objects and measurements. In practice, Bernoulli component pruning limits \(n_{k|k-1}\), while message censoring reduces the number of active association edges \cite{florian2020scalable,florian2021scalable}.

For particle-based implementations, the quality of the proposal density is important. Specifically, for new Bernoulli components, instead of drawing particles from a non-informative proposal, e.g., a uniform density for the locations, one may use a measurement-driven proposal density that depends on the associated measurement or measurement cluster. This is especially useful for state components that are directly informed by the measurements. In addition, systematic resampling is applied when needed in order to alleviate particle degeneracy \cite{arulampalam2002tutorial}.

Lastly, for state estimation, we extract object estimates from Bernoulli components whose existence probabilities exceed a threshold. For each such Bernoulli component, the estimate is obtained from the weighted particle set, e.g., by weighted averaging of the state components.

\section{Simulations and Experiments}
\label{sec_simulation}
In this section, we present the results from a Monte Carlo simulation study with 100 runs where the performance of the following four multiple extended object filters are compared\footnote{PMB-PPP-BP implementation is modified based on MATLAB code at
\url{https://github.com/yuhsuansia/Trajectory-PMB-EOT-BP} to account for an unknown Poisson measurement rate. MATLAB code of PMB(M)-SO-ZIP is available at \url{https://github.com/yuhsuansia/Extended-target-PMBM-tracker}. MATLAB code of PMB-BP-ZIP will be available at \url{https://github.com/yuhsuansia} upon acceptance.}:
\begin{itemize}
  \item The proposed extended object PMB filter under the ZIP measurement model, referred to as PMB-BP-ZIP.
  \item The extended object PMB filter under the Poisson measurement model with BP, referred to as PMB-BP-PPP.
  \item The extended object PMBM filter under the ZIP measurement model with sampling-based data association \cite{soextended}, referred to as PMBM-SO-ZIP.
  \item The extended object PMB filter under the ZIP measurement model with sampling-based data association \cite{soextended}, referred to as PMB-SO-ZIP.
\end{itemize}

We consider the random matrix model for EOT, where the object has an elliptical shape modeled by a symmetric positive definite matrix \cite{randomMatrix2}. The random matrix model has been used in several PMBM and PMB implementations \cite{pmbmextended2,xia2021poisson,soextended,xia2023trajectory}, thereby making the comparison easy. In addition, we also validate the proposed PMB-BP-ZIP filter on real-world lidar data for pedestrian tracking.

\subsection{Simulation setting}

The single object state is $x_k = (e_k, E_k, \gamma_k)$, consisting of the kinematic state $e_k\in \mathbb{R}^4$, a $2\times 2$ symmetric positive definite matrix representing the extent state $E_k$, and the Poisson object measurement rate $\gamma_k> 0$. The kinematic state $e_k$ consists of 2D position and velocity. We assume that each object moves according to a nearly constant velocity motion model, and that its extent and measurement rate remain unchanged over time. The kinematic state transition density is 
\begin{equation}
  g_k(e_k | e_{k-1}) = \mathcal{N}(e_k; F e_{k-1}, Q), \quad
  F = I_2 \otimes \begin{bmatrix}1 & T\\
0 & 1\end{bmatrix}, \quad
Q = \sigma_q^2I_2 \otimes \begin{bmatrix}T^3/3 & T^2/2\\
T^2/2 & T\end{bmatrix},
\end{equation}
where $I_2$ is the $2\times 2$ identity matrix, $T=0.2$ is the sampling interval, and $\sigma_q=0.8$ is the process noise standard deviation. 

\begin{figure}[!t]
    \centering
    \includegraphics[width=0.5\columnwidth]{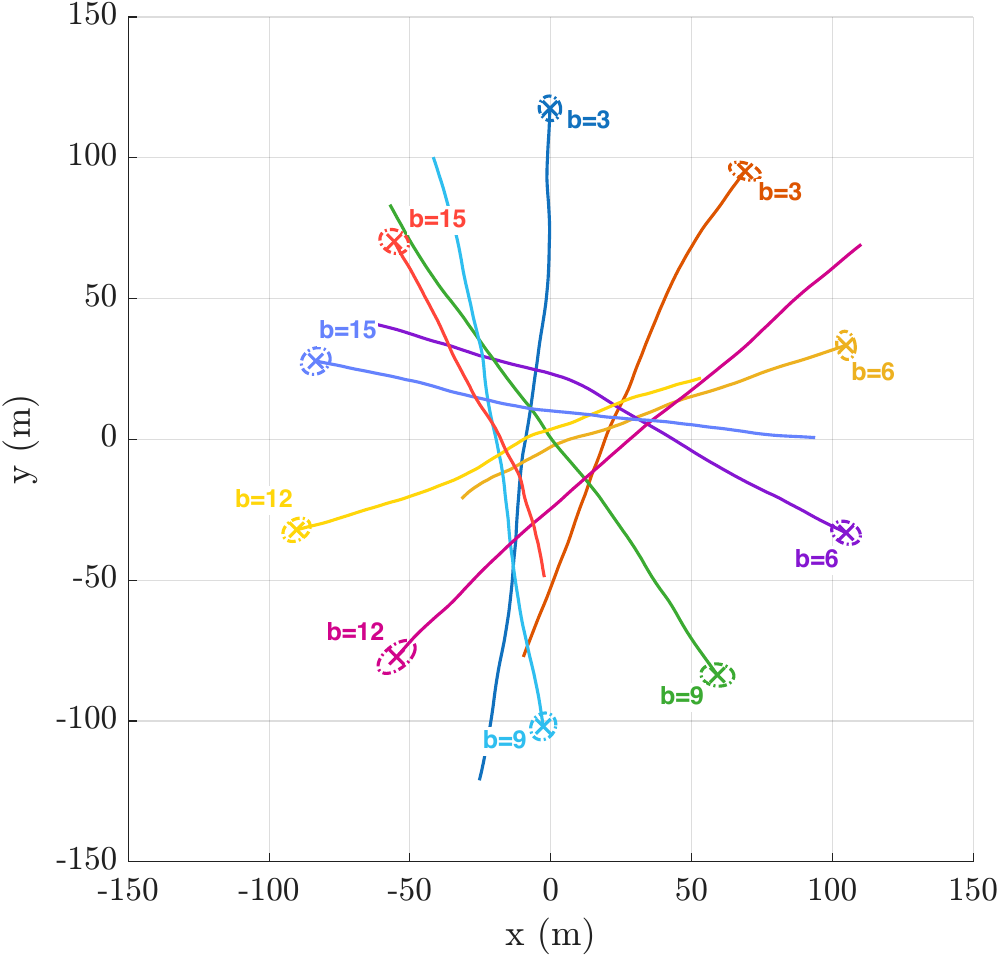}
    \caption{Illustration of the ground truth scenario. Two objects are born at each of the time steps 3, 6, 9, 12, and 15, while two objects are dead at time steps 83, 86, 89, 92, and 95. The initial object locations are indicated by crosses, along with their corresponding elliptical extents. Text annotations specify the birth times; e.g., $b = k$ denotes that an object appears at time step $k$.}
    \label{fig_scenario}
\end{figure}

We consider a similar scenario as in \cite{florian2021scalable,xia2023trajectory} with a region of interest of
size $[-150\,\text{m}, 150\,\text{m}]\times[-150\,\text{m}, 150\,\text{m}]$, where ten objects start moving towards the center from positions uniformly placed on a circle of radius 125\,\text{m} around the center with an initial velocity 12.5\,\text{m}/\text{s}, and then they become closely-spaced for some time before they separate. The initial extent matrix of each object is obtained by sampling from an
inverse-Wishart distribution with degree of freedom 100 and mean matrix $5I_2$. The true object trajectories are illustrated in Fig. \ref{fig_scenario}. The
object survival probability is $p^S = 0.99$.

\begin{figure*}[t]
\centering

\includegraphics[width=0.24\textwidth]{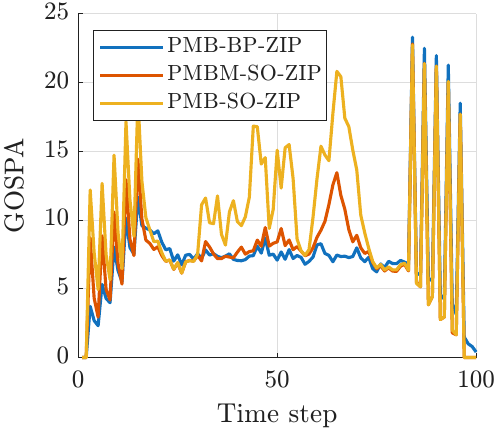}
\includegraphics[width=0.24\textwidth]{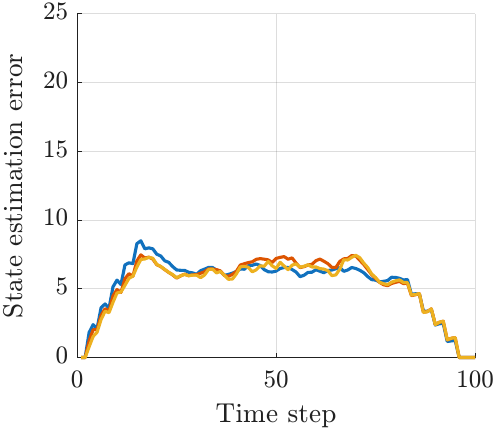}
\includegraphics[width=0.24\textwidth]{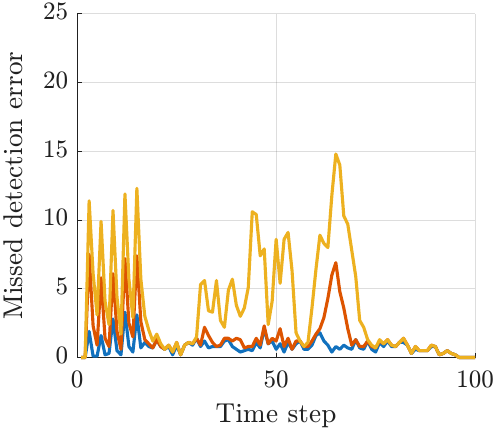}
\includegraphics[width=0.24\textwidth]{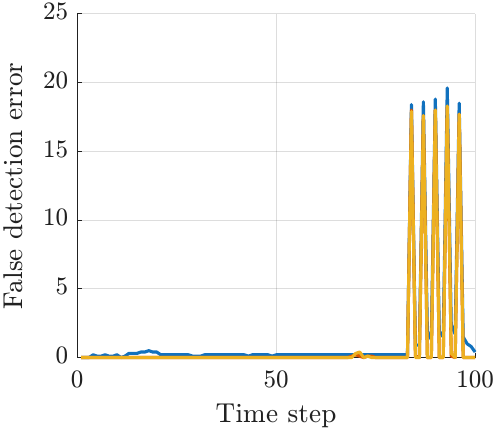}

\par\smallskip
{\footnotesize (a) $p^D = 0.9$, $\gamma = 10$}

\includegraphics[width=0.24\textwidth]{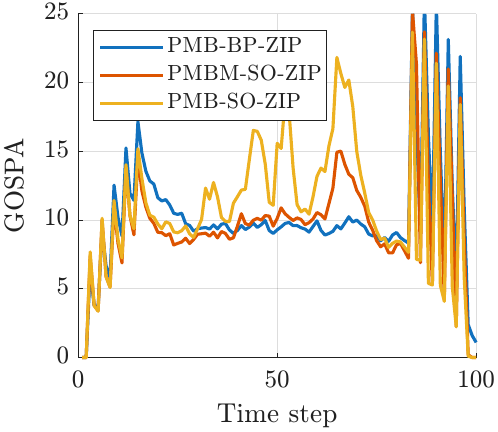}
\includegraphics[width=0.24\textwidth]{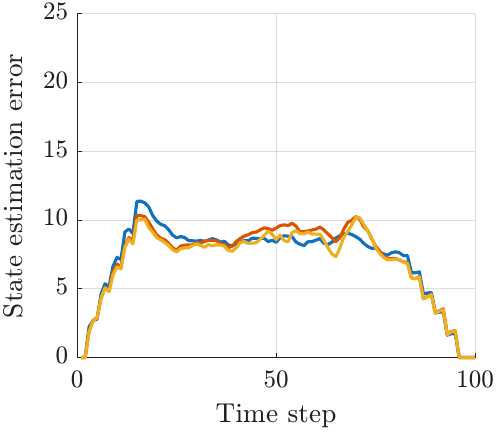}
\includegraphics[width=0.24\textwidth]{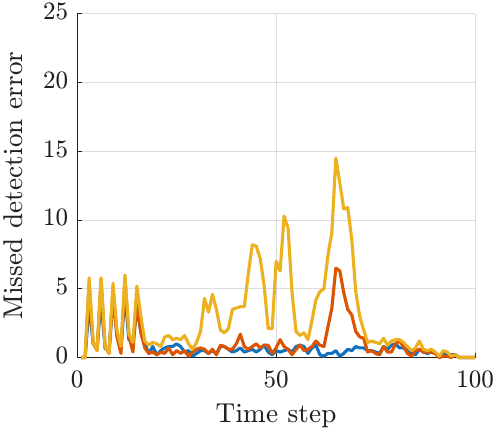}
\includegraphics[width=0.24\textwidth]{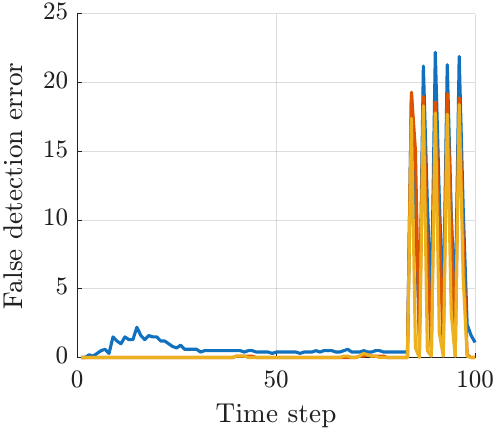}

\par\smallskip
{\footnotesize (b) $p^D = 0.9$, $\gamma = 5$}

\includegraphics[width=0.24\textwidth]{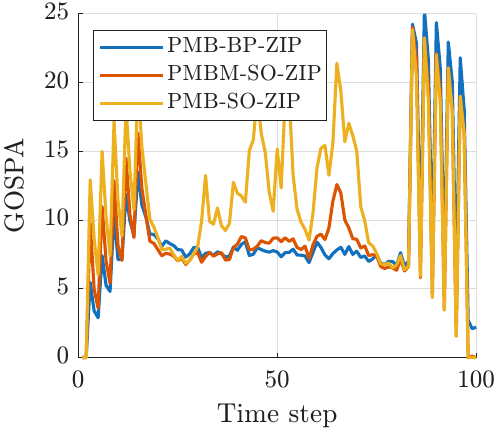}
\includegraphics[width=0.24\textwidth]{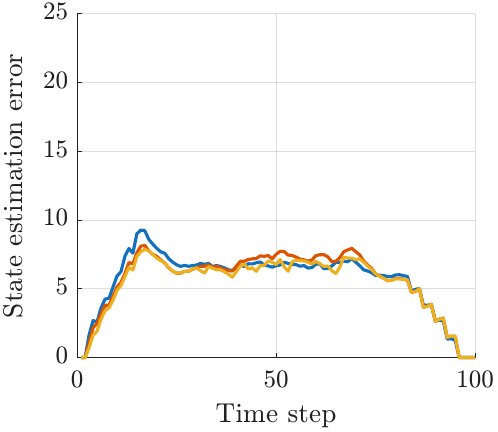}
\includegraphics[width=0.24\textwidth]{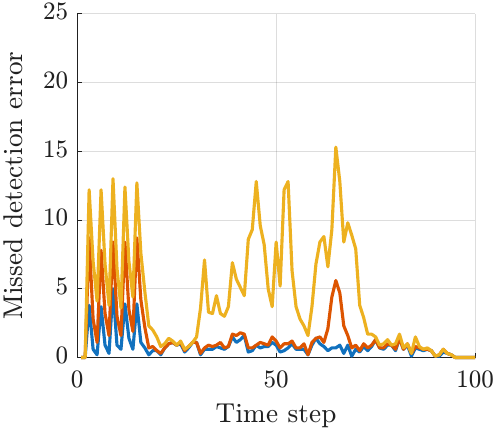}
\includegraphics[width=0.24\textwidth]{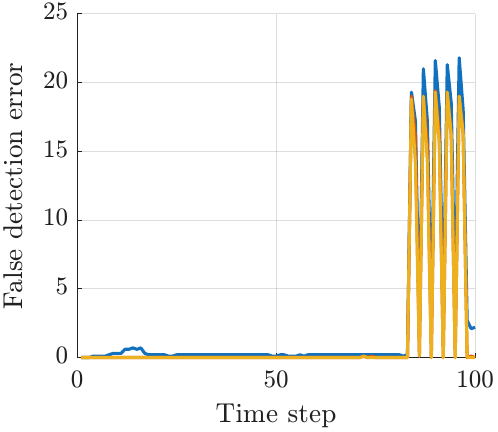}

\par\smallskip
{\footnotesize (c) $p^D = 0.8$, $\gamma = 10$}

\caption{Performance comparison of PMB-BP-ZIP, PMBM-SO-ZIP and PMB-SO-ZIP, showing the GOSPA error and its decomposition over time.}
\label{fig:results}
\end{figure*}

\subsection{Implementation details}

Both PMB-BP-ZIP and PMB-BP-PPP use particle implementations, whereas PMBM-SO-ZIP and PMB-SO-ZIP adopt analytical GGIW implementations \cite{pmbmextended2}. For all the implementations, the Poisson birth rate is 0.01. In the birth density, the velocity is Gaussian distributed with zero mean and covariance $225I_2$, the extent matrix is inverse-Wishart distributed with mean $5I_2$ and degree of freedom 100, and the Poisson object measurement rate is Gamma distributed with mean $\gamma$ and rate parameter 100. For the position, it is Gaussian distributed with zero mean and covariance $150^2I_2$ for GGIW implementations, and it is uniformly distributed in the entire region for particle-based implementations. In addition, for particle-based implementations, the proposal distribution for the kinematic state is given by its motion model, whereas for the extent matrix its proposal is given by a mean-preserving Wishart distribution with degree of freedom 10000, and for the measurement rate its proposal is given by a mean-preserving gamma distribution with rate parameter 10000. The details of the GGIW prediction can be found in \cite{pmbmextended2}.

For the measurement model, the single measurement likelihood is Gaussian with
$\ell_k(z|x_k) = \mathcal{N}(z; H e_k, E_k)$, $H = I_2 \otimes [1~0]$.
The object detection probability is $p^D$, and the Poisson object measurement rate is $\gamma$. In the simulation, we consider three different settings with $(p^D,\gamma)\in\{(0.9,10),(0.9,5),(0.8,10)\}$. The Poisson clutter measurements are uniformly
distributed in the region of interest with Poisson rate $\lambda^C = 10$. Since PMB-BP-PPP assumes $p^D = 1$, for a fair comparison, we set the parameter $\gamma$ used in PMB-BP-PPP to be $p^D\gamma $. Under this setting, the expected number of measurements generated by an object is the same for both PPP and ZIP measurement models. 

For PMB-BP-ZIP and PMB-BP-PPP, we use 5000 particles 5000 with systematic resampling applied at every time step; the number of iterations for loopy BP is 3. In addition, we use message censoring with threshold 0.9 to reduce computational complexity and measurement reordering to facilitate track initialization. We also adopt measurement-driven initialization for newly detected objects \cite{florian2020scalable}. Specifically, for a particle representing a newly detected object, the location is sampled from a Gaussian proposal whose mean is given by the centroid of the corresponding measurement cluster and whose covariance is set to twice the cluster sample covariance, with the measurement clusters obtained via distance partitioning \cite{phdextended2}. The velocity, extent, and Poisson measurement rate are sampled from the birth density.

To ensure computational tractability, we prune Bernoulli components with existence probability smaller than $10^{-3}$ and also mixture components in the Poisson intensity for undetected objects with weight smaller than $10^{-3}$. For PMBM-SO-ZIP and PMB-SO-ZIP, the pruning thresholds for global hypotheses, GGIW components in the Poisson intensity for undetected objects and Bernoulli components are set to $10^{-2}$, $10^{-3}$ and $10^{-3}$, respectively. We also apply ellipsoidal gating with a gate size 13.8. The number of sampling iterations for data association is set to $\lceil100m_kw_a\rceil$ for global hypothesis with weight $w_a$ in PMBM-SO-ZIP and $50m_k$ in PMB-SO-ZIP.

To extract object state estimates from the filtering density, for all the filters, we apply a threshold of 0.5 on the existence probabilities of Bernoulli components. As for PMBM-SO-ZIP, since it maintains multiple global hypotheses, we first select the global hypothesis with the highest weight and then extract estimates from its Bernoulli components. The performance is evaluated using the generalized optimal subpattern assignment (GOSPA) metric \cite{gospa} with $\alpha=2$, exponent $p=1$, cut-off $c=20$. The base measure used in GOSPA is the square root Gaussian Wasserstein distance for comparing two extended object states with elliptical shapes \cite{gwd}. 

\subsection{Simulation results}

\begin{table*}[!t]
  \caption{Multi-object filtering performance evaluated using the GOSPA metric, together with its decomposition into state estimation, missed detection, and false detection errors, as well as runtime (in seconds), averaged over 100 time steps.}
  \centering
\resizebox{\textwidth}{!}{%
\begin{tabular}{lccccccccccccccc}
\toprule
 & \multicolumn{5}{c}{$p^D = 0.9$, $\gamma = 10$} & \multicolumn{5}{c}{$p^D = 0.9$, $\gamma = 5$} & \multicolumn{5}{c}{$p^D = 0.8$, $\gamma = 10$} \\
\cmidrule(lr){2-6}\cmidrule(lr){7-11}\cmidrule(lr){12-16}
 & Total & State & Miss & False & Runtime & Total & State & Miss & False & Runtime & Total & State & Miss & False & Runtime \\
\midrule
PMB-BP-ZIP & \textbf{7.39} & 5.32 & 0.80 & 1.26 & 219.5 & 10.16 & 7.16 & 0.70 & 2.29 & 119.1 & \textbf{8.80} & 5.68 & 0.81 & 2.31 & 197.1 \\
\midrule
PMB-BP-PPP & 63.39 & 10.88 & 6.62 & 45.89 & 374.1 & 13.36 & 7.47 & 3.59 & 2.29 & 118.6 & 62.81 & 12.90 & 12.78 & 37.13 & 325.3 \\
\midrule
PMBM-SO-ZIP & 7.76 & 5.34 & 1.51 & 0.90 & 576.2 & \textbf{9.76} & 7.20 & 1.08 & 1.49 & 324.5 & 8.92 & 5.67 & 1.49 & 1.76 & 544.9 \\
\midrule
PMB-SO-ZIP & 9.90 & 5.19 & 3.81 & 0.91 & 354.3 & 11.94 & 6.95 & 2.96 & 1.03 & 173.1 & 11.46 & 5.41 & 4.33 & 1.72 & 320.1 \\
\bottomrule
\end{tabular}}
  \label{tab_gospa}
\end{table*}

\begin{table*}[!t]
  \caption{An ablation study of the PMB-BP-ZIP filter evaluated using GOSPA, together with its decomposition into state estimation, missed detection, and false detection errors, as well as runtime (in seconds), averaged over 100 time steps. Unless otherwise stated, $N_p=5000$, $N_{\mathrm{BP}}=3$,
  and message censoring and measurement
  reordering are enabled.}
  \centering
  \resizebox{\textwidth}{!}{%
  \begin{tabular}{lccccccccccccccc}
    \toprule
     & \multicolumn{5}{c}{$p^D=0.9$, $\gamma=10$}
     & \multicolumn{5}{c}{$p^D=0.9$, $\gamma=5$}
     & \multicolumn{5}{c}{$p^D=0.8$, $\gamma=10$} \\
    \cmidrule(lr){2-6}
    \cmidrule(lr){7-11}
    \cmidrule(lr){12-16}
     & Total & State & Miss & False & Runtime
     & Total & State & Miss & False & Runtime
     & Total & State & Miss & False & Runtime \\
    \midrule
    $N_{\mathrm{BP}}=2$
      & 7.42 & 5.32 & 0.77 & 1.33 & 149.0
      & 10.71 & 7.14 & 0.70 & 2.87 & 86.5
      & 9.04 & 5.70 & 0.80 & 2.54 & 134.4 \\
    $N_{\mathrm{BP}}=3$ (baseline)
      & 7.39 & 5.32 & 0.80 & 1.26 & 208.9
      & 10.16 & 7.16 & 0.70 & 2.29 & 121.0
      & 8.80 & 5.68 & 0.81 & 2.31 & 194.1 \\
    $N_{\mathrm{BP}}=4$
      & 7.18 & 5.27 & 0.82 & 1.10 & 271.1
      & 9.71 & 7.12 & 0.74 & 1.85 & 153.4
      & 8.60 & 5.61 & 0.83 & 2.16 & 254.8 \\
    $N_{\mathrm{BP}}=5$
      & \textbf{7.17} & 5.27 & 0.82 & 1.09 & 325.7
      & \textbf{9.68} & 7.14 & 0.74 & 1.81 & 187.2
      & \textbf{8.59} & 5.61 & 0.83 & 2.15 & 314.8 \\
    \midrule
    $N_p=2000$
      & 7.67 & 5.57 & 0.82 & 1.28 & 69.7
      & 10.49 & 7.44 & 0.76 & 2.29 & 38.5
      & 9.12 & 5.93 & 0.82 & 2.38 & 67.7 \\
    $N_p=10000$
      & 7.37 & 5.24 & 0.81 & 1.33 & 398.1
      & 10.16 & 7.06 & 0.70 & 2.41 & 253.5
      & 8.62 & 5.57 & 0.82 & 2.23 & 397.9 \\
    \midrule
    w/o measurement reordering
      & 7.67 & 5.36 & 1.07 & 1.23 & 191.4
      & 10.68 & 7.21 & 0.93 & 2.55 & 117.9
      & 9.51 & 5.75 & 1.14 & 2.61 & 197.7 \\
    w/o message censoring
      & 7.25 & 5.32 & 0.81 & 1.12 & 460.5
      & 10.75 & 7.18 & 0.72 & 2.85 & 315.2
      & 8.92 & 5.69 & 0.81 & 2.43 & 468.6 \\
    \bottomrule
  \end{tabular}}
  \label{tab:pmb_bp_zip_ablation}
\end{table*}

Table~\ref{tab_gospa} summarizes the performance of the different filters in terms of GOSPA error, and its decomposition into state estimation, missed detection, and false detection errors, as well as runtime. The results show that PMB-BP-ZIP generally achieves the lowest total GOSPA error, except for $(p^D,\gamma)=(0.9,5)$, for which PMBM-SO-ZIP achieves the lowest value. In particular, PMB-BP-ZIP significantly outperforms PMB-BP-PPP, which performs poorly due to measurement model mismatch, especially in the case of higher Poisson measurement rate. Although setting the Poisson rate in PMB-BP-PPP to $p^D\gamma$ matches the expected measurement count, the PPP model assigns probability $p^De^{-\gamma}$ to an empty measurement set, which can be orders of magnitude smaller than the true ZIP probability $1-p^D+p^De^{-\gamma}$. This discrepancy increases rapidly with $\gamma$, causing missed detection updates to sharply reduce the existence probabilities of existing Bernoulli components. Consequently, objects are more likely to be reinitialized when measurements reappear. Moreover, since marginal Bernoulli densities obtained in BP do not preserve negative correlations among competing local hypotheses for newly detected objects, multiple nearby Bernoulli components may exceed the extraction threshold simultaneously, producing duplicate  estimates and thus a large false detection error.

Compared to PMBM-SO-ZIP and PMB-SO-ZIP, PMB-BP-ZIP has lower GOSPA error in two out of three settings, and it reports a lower missed detection error than both PMBM-SO-ZIP and PMB-SO-ZIP across all three settings. In addition, PMB-BP-ZIP has a lower runtime than both PMBM-SO-ZIP and PMB-SO-ZIP across all three settings. We also note that the differences in GOSPA errors between PMB-BP-ZIP and PMBM-SO-ZIP are less pronounced than those reported in \cite{xia2023trajectory} for a similar scenario but with the Poisson object measurement model. This may be attributed to the fact that, the ZIP measurement model introduces additional cycles in the factor graph, which can degrade the accuracy of the loopy BP approximation. 

To further analyze the filtering performance of PMB-BP-ZIP, PMBM-SO-ZIP, and PMB-SO-ZIP, Fig.~\ref{fig:results} presents the GOSPA error and its decomposition over time for the three scenario settings. It can be observed that PMB-BP-ZIP yields a substantially lower missed detection error than both PMBM-SO-ZIP and PMB-SO-ZIP across all three settings, especially at the time steps when objects are born and when closely-spaced objects become separated. The superior track initialization performance of PMB-BP-ZIP might be attributed to its measurement-driven proposal distribution for new Bernoulli components, although this comes at the cost of increased sensitivity to false detections. We also note that, although both PMB-BP-ZIP and PMB-SO-ZIP employ an MB approximation, only PMB-SO-ZIP suffers from coalescence, while PMB-BP-ZIP performs significantly better when objects are moving in proximity. This advantage stems from its particle-based state representation, which can better capture posterior multi-modality than the parametric representation adopted in PMBM-SO-ZIP and PMB-SO-ZIP.

Having established the comparative performance of the considered filters, we next examine how the implementation choices affect the estimation accuracy and computational cost of PMB-BP-ZIP. As reported in Table~\ref{tab:pmb_bp_zip_ablation}, increasing the number of BP iterations from two to five consistently reduces the GOSPA error, and the performance largely saturates at four. Similarly, increasing the number of particles from $2000$ to $5000$ improves estimation accuracy, whereas using $10000$ particles yields only marginal additional gains at approximately twice the runtime.

The local hypothesis structure for new Bernoulli components depends on the measurement ordering. Specifically, a component initiated by measurement $z_k^j$ can only include measurements whose indices do not exceed $j$, see Lemma \ref{pmbm_update}. The reordering heuristic described in \cite{meyer2020scalable2} places measurements belonging to the same cluster before the corresponding representative measurement. Disabling this reordering increases the GOSPA error in all the three scenarios. Message censoring further reduces complexity by neglecting weak new Bernoulli association branches for measurements that are already sufficiently explained by the existing Bernoulli components. Disabling this censoring doubles the runtime, while producing only modest changes in GOSPA error. Hence, measurement reordering improves estimation accuracy, whereas message censoring provides a substantial computational saving with little overall effect on accuracy.

\subsection{Experiment}

\begin{figure}[!t]
  \centering
  \includegraphics[width=0.7\columnwidth]{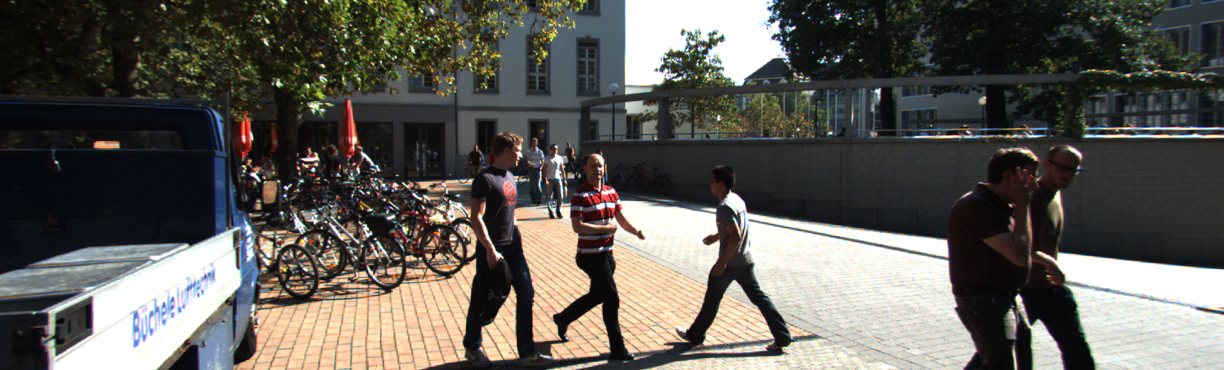}
  \includegraphics[width=0.5\columnwidth]{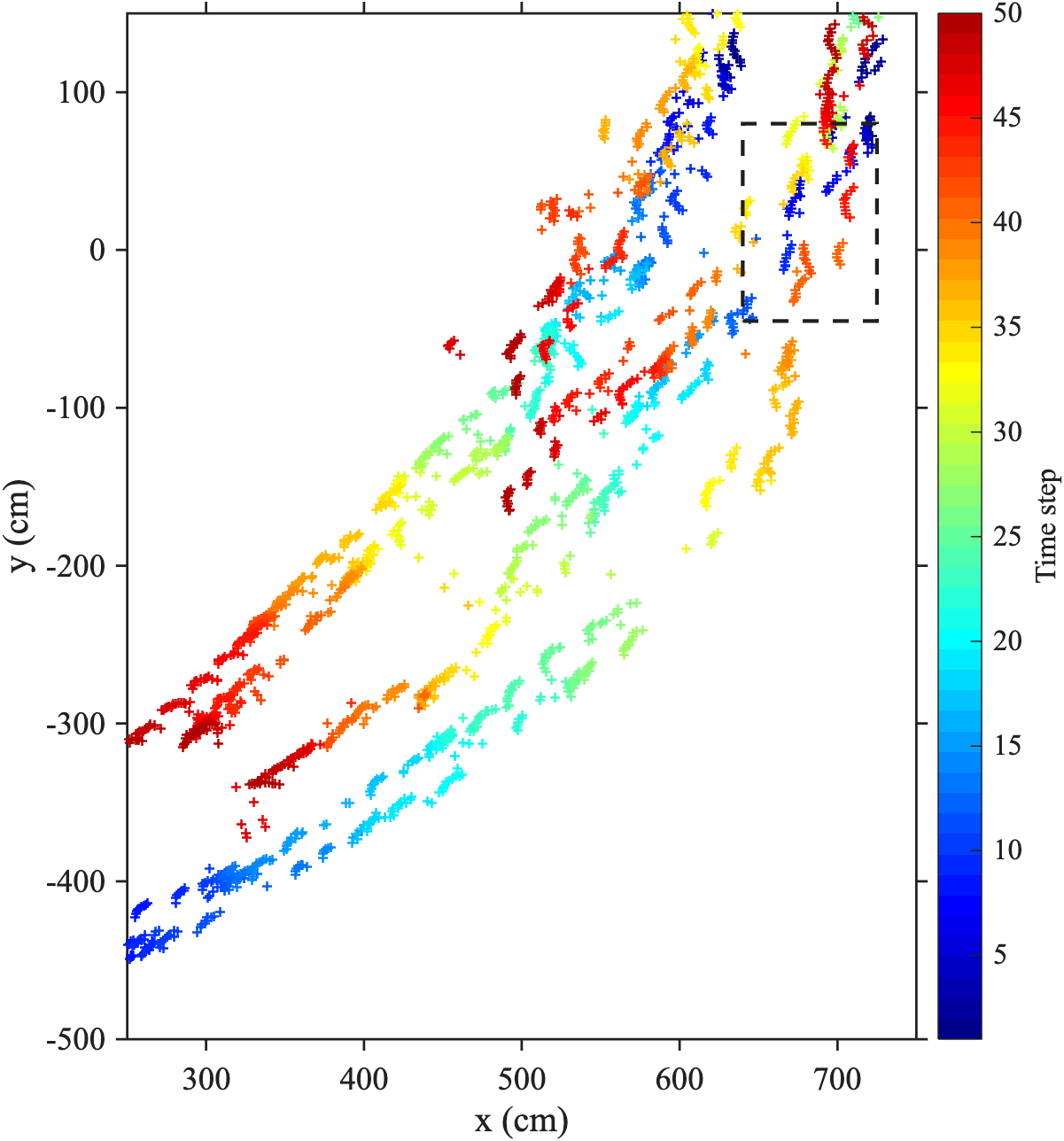}
  \caption{Illustration of real-world camera and LiDAR data from the KITTI sequence \texttt{2011\_09\_28\_drive\_0043}, featuring five walking pedestrians. The top panel shows left-camera image at frame 35, including a closely-spaced central group and an overlapping pair on the right. The bottom panel presents the selected LiDAR measurements, color-coded by time. The dashed rectangle highlights an interaction region containing occlusion-induced measurement gaps.}
  \label{fig_lidar}
\end{figure}

\begin{figure}[!t]
    \centering
    \subfloat[PMB-BP-PPP\label{fig:kitti_ppp}]{%
        \includegraphics[width=0.49\textwidth]
        {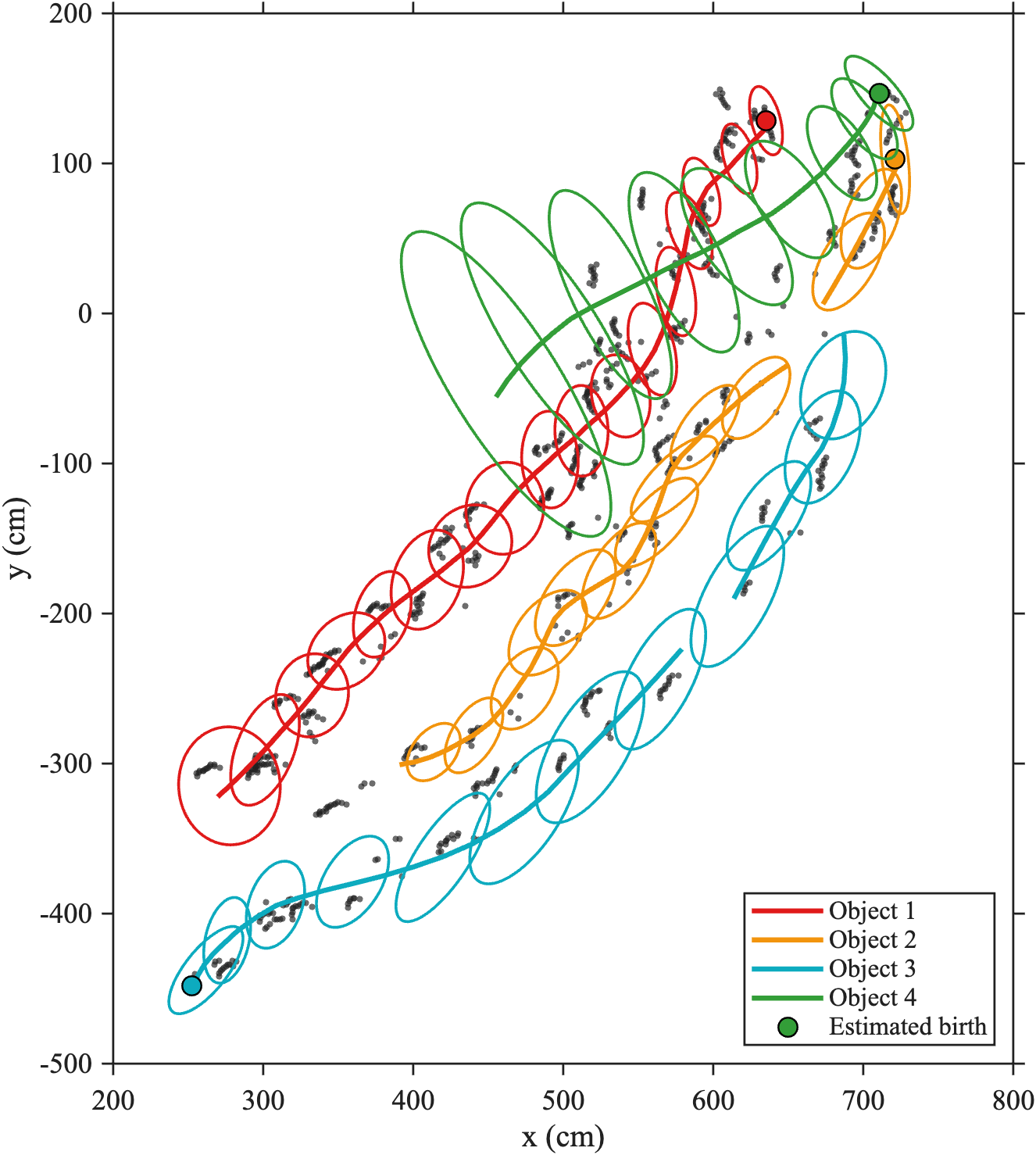}}
    \hfill
    \subfloat[PMB-BP-ZIP\label{fig:kitti_zip}]{%
        \includegraphics[width=0.49\textwidth]
        {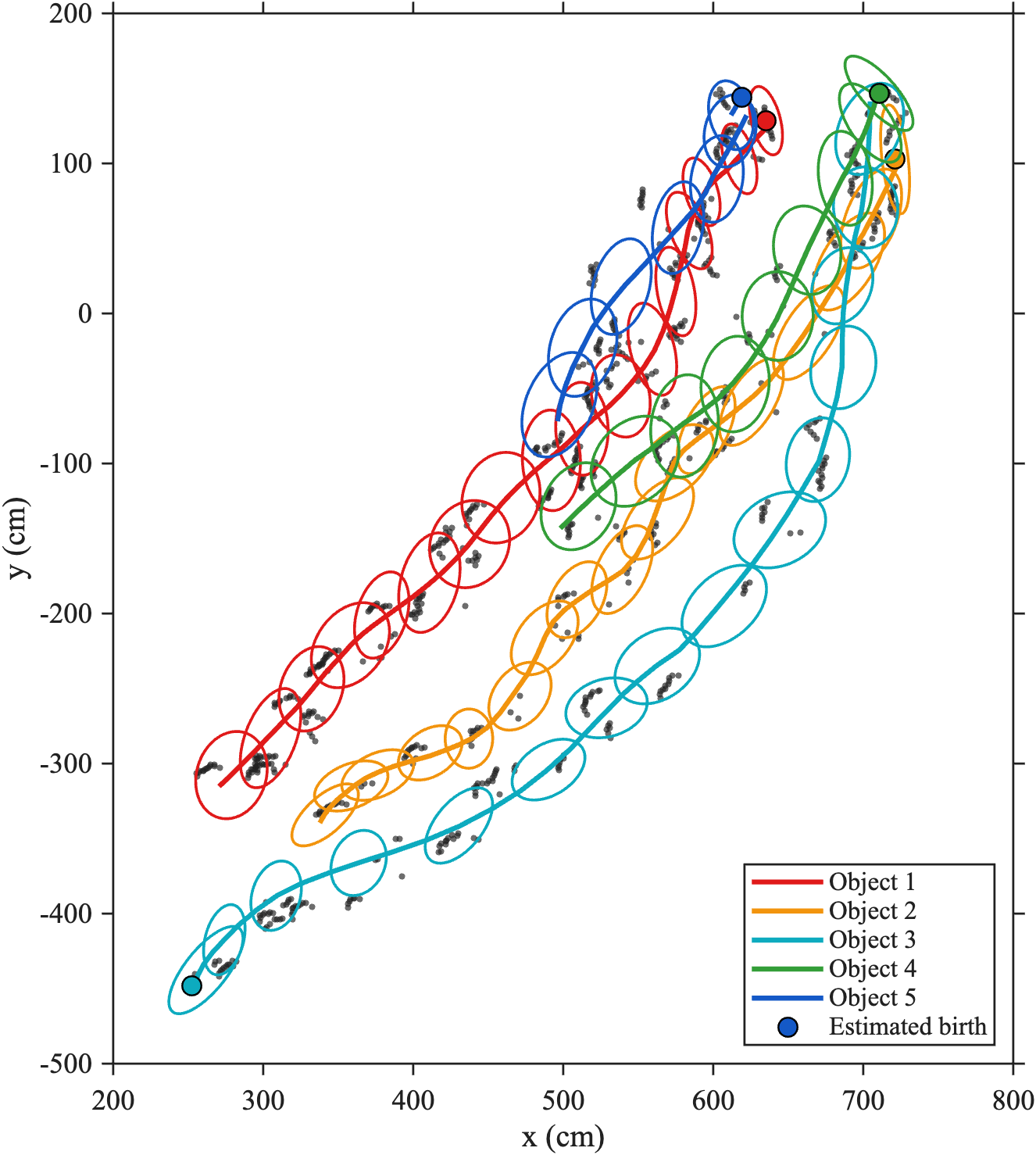}}
    \caption{Pedestrian tracking results obtained using PMB-BP-PPP and PMB-BP-ZIP on a selected KITTI sequence. The estimated object states are represented by ellipses, with the color indicating the track index, illustrated every 3 frames.}
    \label{fig:kitti_tracking_comparison}
\end{figure}

We further evaluate the performance of PMB-BP-ZIP using real-world lidar data from the KITTI dataset~\cite{geiger2013vision}. We consider the first 50 time steps of the tracking sequence \texttt{2011\_09\_28\_drive\_0043}. The sensing platform remains approximately stationary during this interval, and hence ego-motion compensation is unnecessary. The surveillance region is
$[250,750]\,\mathrm{cm}\times[-500,150]\,\mathrm{cm}$. After compensating for the lidar sensor mounting height and road inclination using an estimated ground plane, points between 39 and 43~cm above the road are retained. This slice approximately corresponds to pedestrian knee height and produces measurement clusters that can be represented by ellipses \cite{yang2020marginal}. As shown in Fig.~\ref{fig_lidar}, five pedestrians traverse the region of interest, with several close interactions and complete occlusions between two of them. Since object-level tracking annotations are unavailable for this sequence, we focus on qualitative evaluation.

To isolate the effect of the measurement model, PMB-BP-ZIP and PMB-BP-PPP use identical parameter settings except for the object detection probability $p^D$. Both implementations use 1000 particles, three BP iterations, survival probability $p^{S}=0.99$, Bernoulli extraction threshold $0.5$, Bernoulli pruning threshold $0.01$, and a Poisson clutter rate of one measurement per scan. For PMB-BP-ZIP, the detection probability and conditional measurement rate are set to $p^{D}=0.8$ and $\gamma=8.4$, respectively. PMB-BP-PPP uses $p^{D}=1$ and the same Poisson rate $\gamma=8.4$. Although setting the PPP rate to $p^{D}\gamma$ would match the unconditional expected measurement count, using $\gamma=8.4$ provides better empirical performance in this sequence because complete occlusions occur during only a few time steps. For both filters, the object trajectories are constructed by connecting the estimated object states with the same track index over time.

The qualitative results in Fig.~\ref{fig:kitti_tracking_comparison} demonstrate the importance of explicitly modeling object missed detections. PMB-BP-ZIP maintains all five pedestrian trajectories throughout their visible intervals, with estimated extents generally matching the corresponding lidar returns. In contrast, PMB-BP-PPP produces premature track terminations, missed trajectories, and several excessively large extent estimates during close interactions. This is because the PPP model assigns very low probability to an empty object-generated measurement set. Temporary occlusions therefore provide excessive evidence against object existence. Once a Bernoulli component has small existence probability or is pruned, nearby returns may instead be associated with another component, resulting in distorted trajectories and inflated extent estimates. By explicitly accounting for complete missed detections, PMB-BP-ZIP preserves object existence through temporary measurement gaps and provides more consistent tracking under occlusion.

\section{Conclusions}
\label{sec_conclusion}
In this paper, we have presented an extended object PMB filter under the ZIP object measurement model and its implementation using particle BP. This is derived by applying loopy BP to the factor graph representation of the joint factorized posterior over set of objects, object detection variables, and dual representation of measurement-oriented and object-oriented data association variables. The results show that the proposed filter achieves filtering performance comparable to the sampling-based PMBM implementation, while having a lower runtime. The efficacy of the proposed filter has also been validated on real-world lidar data for pedestrian tracking.

Potential future research directions include the extension to sets of trajectories to enable principled object trajectory estimation \cite{garcia2020trajectory,granstrom2024poisson,xia2023trajectory}, the development of BP-based implementations for more general object and clutter measurement models \cite{garcia2021poisson,garcia2023poisson}, the combination with mean-field variational inference to further reduce the computational complexity \cite{ma2026closed}, and the extension to multi-sensor fusion and multi-scan smoothing \cite{williams2018multiple} using set-type BP \cite{kim2024set}.

\appendix
\section{Algebraic consistency of the factorization in
\cite{lyu2026fast}}
\label{app:factorization_consistency}

This appendix examines the legacy-object part of the joint posterior factorization in \cite[Eq.~10]{lyu2026fast}, whose derivation is given in Eqs.~(B.18)--(B.22) of its supplementary material. Specifically, we verify whether the legacy object factors in Eqs.~(B.19) and (B.22) reproduce the corresponding branches of the joint posterior.

We follow the notation of \cite{lyu2026fast}. For a legacy label $l_s$, let
\begin{align}
\mathcal I_s(\mathbf b_k)
  &\triangleq \{i:b_k^i=l_s\}, \\
n_s(\mathbf b_k)
  &\triangleq |\mathcal I_s(\mathbf b_k)|,\\
\alpha_s(x)
  &\triangleq P_D(x,l_s)e^{-\gamma_k(x,l_s)},\\ 
q_{0,s}(x)
  &\triangleq 1-P_D(x,l_s)+\alpha_s(x),\\
\ell_{s,i}(x)
  &\triangleq
  \frac{\gamma_k(x,l_s)\phi_k(z_k^i | x,l_s)}
       {\lambda_k^C(z_k^i)} .
\end{align}
Here, $q_{0,s}(x)$ is the probability that an existing object produces an empty measurement set.

Using the same clutter normalization as in Eq.~(B.22), the legacy-object contribution specified by Eq.~(B.16) is
\begin{equation}
\Phi_s(\dot{\mathbf X}_k^{l_s},\mathbf b_k)
=
\begin{cases}
1-r_s,
  & \dot{\mathbf X}_k^{l_s}=\emptyset,\ n_s=0,\\[1mm]
r_s p_s(x)q_{0,s}(x),
  & \dot{\mathbf X}_k^{l_s}=\{(x,l_s)\},\ n_s=0,\\[1mm]
r_s p_s(x)\alpha_s(x)
\displaystyle\prod_{i\in\mathcal I_s(\mathbf b_k)}
\ell_{s,i}(x),
  & \dot{\mathbf X}_k^{l_s}=\{(x,l_s)\},\ n_s>0,\\[2mm]
0,
  & \dot{\mathbf X}_k^{l_s}=\emptyset,\ n_s>0,
\end{cases}
\label{eq:legacy_exact}
\end{equation}
where $r_s\triangleq r_{k|k-1}^{l_s}$ and $p_s\triangleq p_{k|k-1}^{l_s}$.

Let $\widehat{\Phi}_s$ denote the product of the factors printed in Eqs.~(B.19) and (B.22):
\begin{equation}
\widehat{\Phi}_s
\triangleq
f_s^{(\mathrm{B.19})}(\dot{\mathbf X}_k^{l_s})
\prod_i
h_{s,i}^{(\mathrm{B.22})}
(\dot{\mathbf X}_k^{l_s},b_k^i;z_k^i).
\end{equation}
For a present legacy object, these factors give
\begin{equation}
\widehat{\Phi}_s
\bigl(\{(x,l_s)\},\mathbf b_k\bigr)
=
r_s p_s(x)\alpha_s(x)
\prod_{i\in\mathcal I_s(\mathbf b_k)}\ell_{s,i}(x).
\label{eq:legacy_printed}
\end{equation}
Consequently, on the nonzero present-object branches,
\begin{equation}
\frac{\widehat{\Phi}_s}{\Phi_s}
=
\begin{cases}
\dfrac{\alpha_s(x)}{q_{0,s}(x)},
  & n_s(\mathbf b_k)=0,\\[3mm]
1,
  & n_s(\mathbf b_k)>0.
\end{cases}
\label{eq:branch_ratios}
\end{equation}

For two expressions to define the same joint posterior up to normalization, their ratio must be independent of both the object state and the association hypothesis. Eq.~\eqref{eq:branch_ratios} violates this requirement whenever both association branches are admissible and
\begin{equation}
\frac{P_D(x,l_s)e^{-\gamma_k(x,l_s)}}
     {1-P_D(x,l_s)+P_D(x,l_s)e^{-\gamma_k(x,l_s)}}
\neq 1.
\end{equation}
Thus, the discrepancy cannot be absorbed into the global normalization constant. Equality holds only when $P_D(x,l_s)=1$ throughout the relevant state support. Therefore, Eqs.~(B.19) and (B.22), as printed, do not in general reproduce the legacy-object contribution of the preceding joint posterior.

A corrected joint posterior factorization is obtained by replacing each legacy-object bracket in Eq.~(B.18) by the factor
$\Phi_s$ in Eq.~\eqref{eq:legacy_exact}. In particular, collecting the unchanged distinct-label, Poisson birth, clutter, and newborn object terms in $\mathcal R_k$, the corrected expression has the form
\begin{equation}
f_k(\dot{\mathbf X}_k,\mathbf b_k | Z_{1:k})
\propto
\mathcal R_k(\dot{\mathbf X}_k,\mathbf b_k)
\prod_{s\in\mathcal S_{\mathrm{legacy}}}
\Phi_s(\dot{\mathbf X}_k^{l_s},\mathbf b_k).
\label{eq:corrected_joint_posterior}
\end{equation}
The factor $\Phi_s$ depends jointly on all association variables through the event $n_s(\mathbf b_k)=0$. Hence, an exact pairwise factor graph must represent this event using an auxiliary object-level variable, or an equivalent construction, rather than only the factors in Eqs.~(B.19) and (B.22).

If $P_D$ and $\gamma_k$ are instead interpreted as state-independent scalars, the printed factors can equivalently be corrected by adding
\begin{equation}
\kappa_s(\mathbf b_k)
=
\begin{cases}
\dfrac{1-P_D+P_De^{-\gamma_k}}
      {P_De^{-\gamma_k}},
  & n_s(\mathbf b_k)=0,\\[3mm]
1,
  & n_s(\mathbf b_k)>0,
\end{cases}
\end{equation}
provided that $P_De^{-\gamma_k}>0$. This scalar special case gives $\Phi_s=f_s^{(\mathrm{B.19})}\kappa_s\prod_i h_{s,i}^{(\mathrm{B.22})}$. The higher-order expression in Eq.~\eqref{eq:legacy_exact}, however, also covers the state-dependent model introduced in \cite[Eq.~3]{lyu2026fast}.

\bibliographystyle{IEEEtran}
\bibliography{mybibli.bib}

\end{document}